 \documentclass[aps, twocolumn, nopacs, amsmath]{revtex4}
\usepackage{dcolumn}
\usepackage{bm}
\usepackage{graphicx}
\usepackage{color}
\usepackage{subfigure}
\usepackage{hyperref}
\usepackage{latexsym}
\usepackage{amsthm}
\usepackage{amssymb}
\usepackage{array}
\usepackage{braket}
\DeclareGraphicsExtensions{.jpg,.pdf, .mps, .png, .eps, .ps, .EPS,.gif}

\begin{document}
\def\be{\begin{equation}}
\def\ee{\end{equation}}

\def\bc{\begin{center}}
\def\ec{\end{center}}
\def\bea{\begin{eqnarray}}
\def\eea{\end{eqnarray}}
\newcommand{\avg}[1]{\langle{#1}\rangle}
\newcommand{\Avg}[1]{\left\langle{#1}\right\rangle}

\def\ie{\textit{i.e.}}
\def\etal{\textit{et al.}}
\def\m{\vec{m}}
\def\G{\mathcal{G}}

\newcommand{\gin}[1]{{\bf\color{blue}#1}}
\newcommand{\filrad}[1]{{\color{red}#1}}
\newcommand{\filradcomm}[1]{{\bf\color{green}#1}}

\title{Epidemic plateau in critical SIR dynamics with non-trivial initial conditions}
\author{Filippo Radicchi}
\address{Center for Complex Networks and Systems Research, Luddy School
  of Informatics, Computing, and Engineering, Indiana University, Bloomington,
  IN 47408}
\author{Ginestra Bianconi}
\affiliation{The Alan Turing Institute, 96 Euston Rd, London NW1 2DB, United Kingdom\\ 
School of Mathematical Sciences, Queen Mary University of London, London, E1 4NS, United Kingdom}

\begin{abstract}
  Containment measures implemented by some countries to suppress the spread of
  COVID-19 have resulted in a slowdown of the epidemic characterized
  by time series of daily infections plateauing over extended periods
  of time.  We prove that such a dynamical pattern is compatible with
  critical Susceptible-Infected-Removed (SIR) dynamics. In
  traditional analyses of the critical SIR
  model, the critical dynamical regime is started from a single
  infected node. The application of containment measures to an
  ongoing epidemic, however, has the effect to make the system enter in its critical regime with a
  number of infected individuals potentially large. We
  describe how such non-trivial starting conditions affect the critical behavior of the SIR model. We
  perform a theoretical and large-scale numerical investigation of the
  model. We show that the expected outbreak
  size is an increasing function of the initial number of infected individuals,
  while the expected duration of the outbreak
  is a non-monotonic function of the initial number of infected individuals. Also,
  we precisely characterize the magnitude of the fluctuations
  associated with the size
  and duration of the outbreak in critical SIR dynamics with
  non-trivial initial conditions. Far from heard immunity, fluctuations are much larger than
  average values, thus indicating that predictions of plateauing
  time series may be particularly challenging.
\end{abstract}

\maketitle
\section{Introduction}

At the onset of the COVID-19 pandemic, world-wide time series
of the
number of infected individuals have displayed an exponential growth. Such a
behavior is well predicted by standard epidemic frameworks~\cite{anderson1992infectious}.
In slightly later stages, however, time series have exhibited
non-trivial dynamical patterns. Many papers have attempted to
model observed behaviors and to determine the role of containment measures~\cite{Nigel,Nigel2,ferretti,bianconi2020message,fanelli,timoteo,bianconi,bradde,Reka,Arenas,
  Ziff,bianconi2020epidemics,Nekovee,Nechaev,Blasius,quadratic}. The
common and reasonable assumption is that containment measures
implemented in the attempt of mitigating the outbreak have strongly
influenced the unfolding of the epidemic. Unfortunately, this a
setting where modeling attempts are particularly challenging. The effective
implementation of containment measures imposed by authorities rely on
people's personal judgements and adaptive behavior, and while epidemic
spreading is a well-studied branch of mathematical
biology~\cite{murray2007mathematical},  statistical
physics~\cite{krapivsky2010kinetic}  and network
science~\cite{barabasi2016network,newman2018networks,bianconi2018multilayer,barrat2008dynamical,dorogovtsev2010lectures,pastor2015epidemic,porter2016dynamical},
the modelling of adaptive behavior
is only at its infancy~\cite{nanni2020give,gross1,gross2}.

According to the data, in several countries, the slowdown of the
epidemic spread is characterized by an almost flat time series of
daily number of new infections.
Moreover, the time series of the number of removed individuals display
power-law
growth instead of an exponential growth as a function of time~\cite{Ziff,Nekovee,Blasius,quadratic}.
Here, we propose a theoretical interpretation of those features as the
signature of the system being in (or near) its critical
regime. Criticality is a fundamental property characterizing the
dynamics of biological
and socio-technical systems~\cite{Bialek,Munoz,gleeson2017temporal}. 
Our work consists of an in-depth investigation of a critical
Susceptible-Infected-Removed (SIR) 
dynamics starting from a non-trivial initial configuration
characterized by $n_0$ initially infected individuals. We interpret
the emergence of the
critical regime as the result of disease containment strategies, and
the non-trivial initial condition as the configuration of the system when
spreading becomes critical. In the typical setting considered in
statistical mechanics~\cite{uno, due,tome2010critical}, a single seed is generally used
as the initial condition for critical  SIR dynamics; the mapping of
the critical SIR to the critical standard branching process allows for
a full characterization of the spreading
dynamics~\cite{zapperi1,zapperi2}.
The realistic assumption of having an initial number of infected
individuals $n_0>1$
introduces an additional scale in the system affecting in a non-trivial manner the scaling
properties of the SIR critical dynamics. While in other non-equilibrium systems
a non-trivial initial condition may lead to a change of the critical
exponent
values~\cite{Hinrichsen,slip_exponent,hinrichsen1998correlated}, in
the critical SIR, the introduction of a non-trivial initial condition
does not change the critical exponents that characterize
the distribution of outbreak size and duration. However, it introduces lower exponential
cutoffs in the distributions. As a result, the expected size and
duration of the outbreak, as
well as their standard deviations, have a non-trivial dependence on the initial condition $n_0$.
In this paper, we evaluate, by means of analytic arguments and
large-scale simulations, the scaling of these quantities
as functions of the population size $N$. 

The paper is structured as follows: in Section II, we provide the theoretical interpretation of the plateau as a critical SIR dynamics starting from $n_0>1$ initial condition; in Section III,   we perform a statistical mechanics investigation of the statistical properties of the critical SIR dynamics with non-trivial initial conditions, supported by extensive numerical simulations of the process; finally, in Section IV, we provide concluding remarks. 
The Appendix describes the Gillespie algorithm used in this work to simulate the critical SIR dynamics. 

\section{The theoretical interpretation of the plateau}

We consider  the Susceptible-Infected-Removed (SIR) model on a well-mixed population~\cite{krapivsky2010kinetic,barabasi2016network,newman2018networks,bianconi2018multilayer,barrat2008dynamical,dorogovtsev2010lectures,pastor2015epidemic,porter2016dynamical}. At any point in time, individuals can be found in three possible states: susceptible, infected and removed. Susceptible individuals do not carry the disease but they can be infected; infected individuals carry the disease, and they can spread it to susceptible individuals; removed individuals are either removed or deceased, and they do no participate in the spreading dynamics. We indicate with $\lambda$ the rate of infection, i.e., the expected number of spreading events occurring per unit of time. Without loss of generality, we set the recovery rate equal to one.

We start our discussion by focusing on the deterministic treatment of
the SIR model on a well-mixed population with infinite size. If we
indicate with $s$, $i$ and $r$ the fractions of susceptible, infected
and removed individuals,
respectively, we can write
\bea
\frac{ds}{dt}&=&-\lambda si,\nonumber \\
\frac{di}{dt}&=&\lambda si-i,\nonumber \\
\frac{dr}{dt}&=&i.
\label{sys}
\eea
Please note that $s+i+r=1$. The critical dynamical regime
is characterized by
\bea
\lambda_c = 1 \; .
\eea
If we start from an initial condition consisting of a fraction $i(0)$ of infected individuals and a fraction $r(0)=0$ of removed individuals, at the onset of the epidemic, i.e., $t \ll 1$,
we observe a different behavior depending on the value of
$\lambda$. In the non-critical regime,  i.e., $\lambda \neq \lambda_c$, the deterministic equations for $i$  and $r$ read 
\bea
\frac{di}{dt}&=&\lambda si-i\simeq (\lambda-1)i,\nonumber \\
\frac{dr}{dt}&=&i.
\eea
Solutions of the above equations  are
\bea
i(t)=i(0)e^{(\lambda-1)t},\nonumber \\
r(t)=\frac{i(0)}{\lambda-1}e^{(\lambda-1)t}.
\eea
In essence, in the subcritical regime, i.e., $\lambda < \lambda_c$, the
number of infected individuals decays exponentially fast, and the
number of removed individuals remains vanishing. In the
supercritical regime, i.e., $\lambda > \lambda_c$, the number of infected and
removed individuals displays an exponential increase.
At criticality, i.e., $\lambda =\lambda_c$, the deterministic equations for $i$ and $r$ are
\bea
\frac{di}{dt}&=& (s-1)i\ll 1,\nonumber \\
\frac{dr}{dt}&=&i \; ,
\eea
leading to
\bea
i(t)\simeq i(0),\nonumber \\
r(t)\simeq i(0) t.
\eea
Therefore, according to the deterministic approach, for small times we should expect that the number of removed individuals at criticality increases linearly in time with a slope that is given by the initial condition $i(0)$,  at the onset of the epidemics.

	\begin{figure}[h!]
	\begin{center}
 \includegraphics[width=0.98\columnwidth]{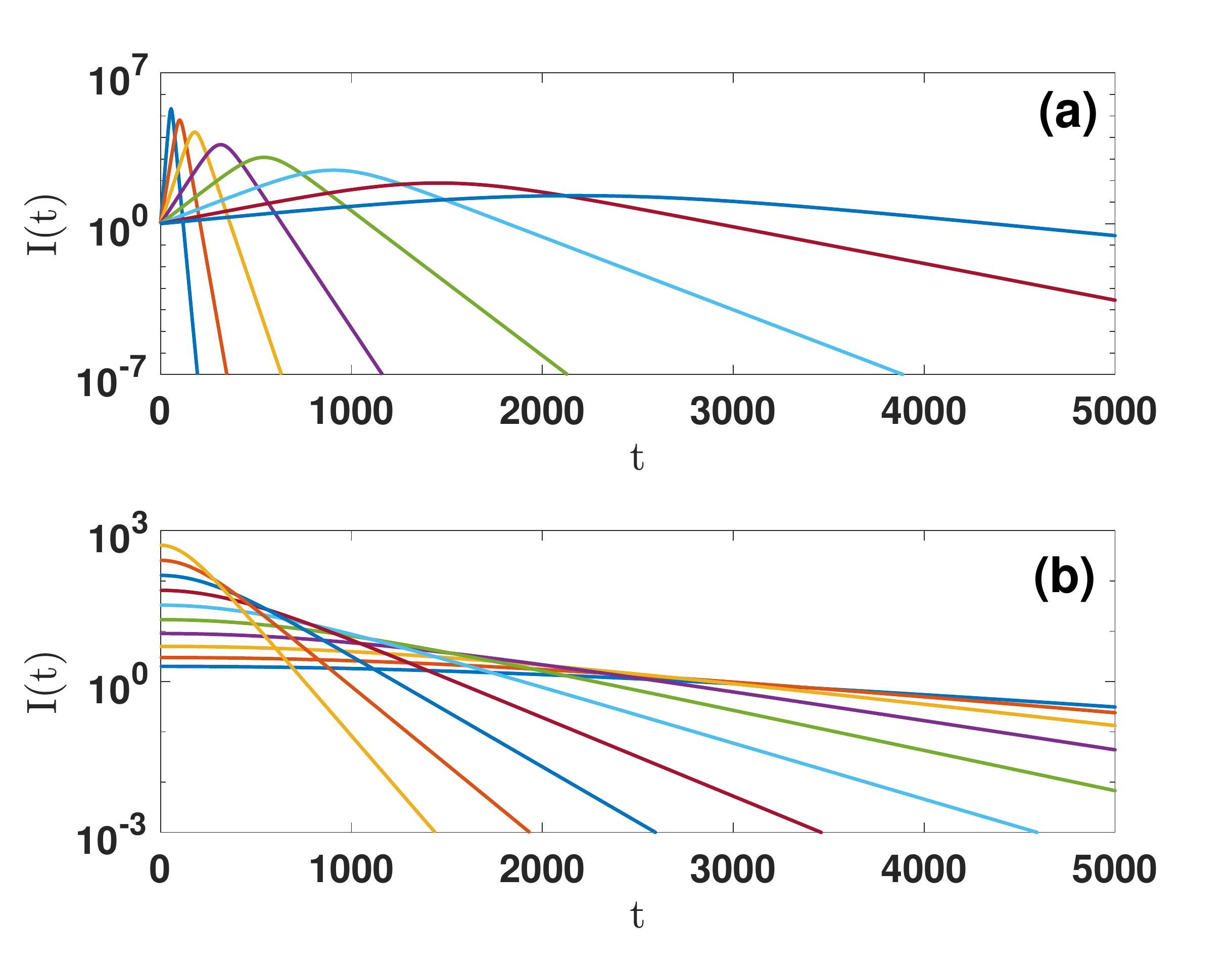}	
\end{center}
  \caption{{a) Time series of  the number of infected individuals
    $I(t) = i(t) \, N$  are plotted close to the critical point. $I(t)$ correspond
    to the solution of the deterministic SIR equations.  The population size is $N=10^7$,
    and spreading is started from $n_0=1$ seed.
    Different curves correspond to different values of
    $\lambda=1+2^{-q}$, with $q = 2, 3, 4, \ldots, 8, 9$.
    b) Same as in panel a,
    but for $\lambda=1$. Different curves correspond to different
    numbers of initially infected nodes $n_0=2^{q}$,
    with $q = 2, 3, 4, \ldots, 8, 9$.
    As $\lambda$ approaches the critical value $\lambda_c=1$ and
    $n_0$ decreases toward one, we observe a plateauing of the time series.}}
  	\label{fig:plateau}
      \end{figure}

From the deterministic Eqs. (\ref{sys}), it  is evident that 
\bea
\frac{di}{ds}=-1+\frac{1}{\lambda s}.
\eea
The equation  can be integrated to obtain the well-known solution   \cite{krapivsky2010kinetic}
\bea
s(t)+i(t)-\frac{1}{\lambda}\ln s(t)=s(0)+i(0)-\frac{1}{\lambda}\ln s(0).
\label{int}
\eea

Using Eqs.(\ref{sys}), we can express the logarithmic derivative of the number of infected individuals as 
\bea
\frac{d\ln i}{dt}=\lambda s-1 , 
\eea
where $\lambda \, s(t)$ is the reproduction number.

The former equation implies that the time series of infected individuals $i(t)$ has a peak at $t=t^{\star}$ determined by
\bea
\lambda s(t^{\star})=1 .
\label{sstar}
\eea
The fraction of susceptible individuals at the peak of the epidemic
is given by $s^{\star}=s(t^{\star})=1/\lambda$. By making the further
assumption that the epidemic starts from a fraction $i(0)=1-s(0)$ of
infected individuals and zero removed individuals $r(0)=0$ in Eq. (\ref{int}), we obtain 
\bea
i^{\star}=i(t^{\star})=1-\frac{1}{\lambda}-\frac{1}{\lambda}\ln(\lambda s(0)).
\label{istar}
\eea
Using Eqs.~(\ref{sstar}) and~(\ref{istar}) in the first of Eqs.~(\ref{sys}), we get
\bea
\left.\frac{ds}{dt}\right|_{t=t^{\star}}=-\lambda s^{\star}i^{\star}=-i^{\star}
\eea
It follows that the second derivative of $\ln i$ is given by   
\bea
\left.\frac{d^2 i}{dt^2}\right|_{t=t^{\star}}
=
i(t^{\star})\left.\frac{d^2 \ln i}{dt^2}\right|_{t=t^{\star}}
= -\lambda (i^{\star})^2=-\rho^{\star},
\label{dR}
\eea
where $\rho^{\star}$ is defined as
\bea
\rho^{\star}=-\frac{1}{\lambda}\left(\lambda-1-\ln (\lambda s(0))\right)^2.
\label{rho}
\eea
We note that $\rho^{\star}$ is zero, i.e., we reach a plateau, only
for $s(0)=1$ and $\lambda=1$. This fact
implies that, in the deterministic approach, a perfect plateau of the
time series $\ln i$ is never achieved for $\lambda>1$.

In the vicinity of the critical point, the time series of
the infected individuals is still well described by a plateau.
Developing the right-hand side of Eq. (\ref{dR}) around $\lambda=1$,
$s(0)=1$,
we get $s(0)\simeq 1$ and 
\bea
\rho^{\star}&\simeq &-\frac{1}{\lambda}\left[\ln(s(0))+\frac{1}{2} (\lambda-1)^2\right]^2\nonumber \\
&\simeq &\left[ 1-s(0)+\frac{1}{2} (\lambda-1)^2\right]^2 .
\eea
The above equation indicates that the conditions to have a near-plateau
dynamics are
having an infectivity rate $\lambda$ close to one, and 
having the system as far as possible from heard immunity, i.e.,
$1-s(0)\ll 1$.
In summary, the near-critical state for $\lambda\simeq 1$ is a fragile state that
can be characterized
by a very slow dynamics if containment measures do not further
decrease the infectivity
$\lambda$ below one  (see Figure~\ref{fig:plateau}).

      \section{SIR critical dynamics with non-trivial initial
        condition}

      From now on, we assume that the system is in the critical
      regime. We further assume that spreading dynamics is started from
      $n_0>1$ initial seeds. The two assumptions serve to rationalize
      two main features of real time series. First, time series
      are characterized by long temporal windows of almost flat
      behavior. This is a signature of criticality.
      Second, plateaus are observed only after initial growths in
      the number of infected individuals, meaning
      that the critical regime is reached only after that containment strategies
      have effectively changed the spreading dynamics of the disease.
      Whereas critical properties of the SIR model are well understood
      for spreading processes initiated by $n_0=1$ individual, we are
      not aware of existing studies dealing with non-trivial initial
      conditions consisting of $n_0 > 1$ seeds.
     How do the properties of the critical dynamics change with
    $n_0$? What is the behavior of the expected duration of the outbreak?
    What about the expected size of the outbreak? What about their fluctuations?

\begin{figure}[h!]
	\begin{center}
          \includegraphics[width=1.0\columnwidth]{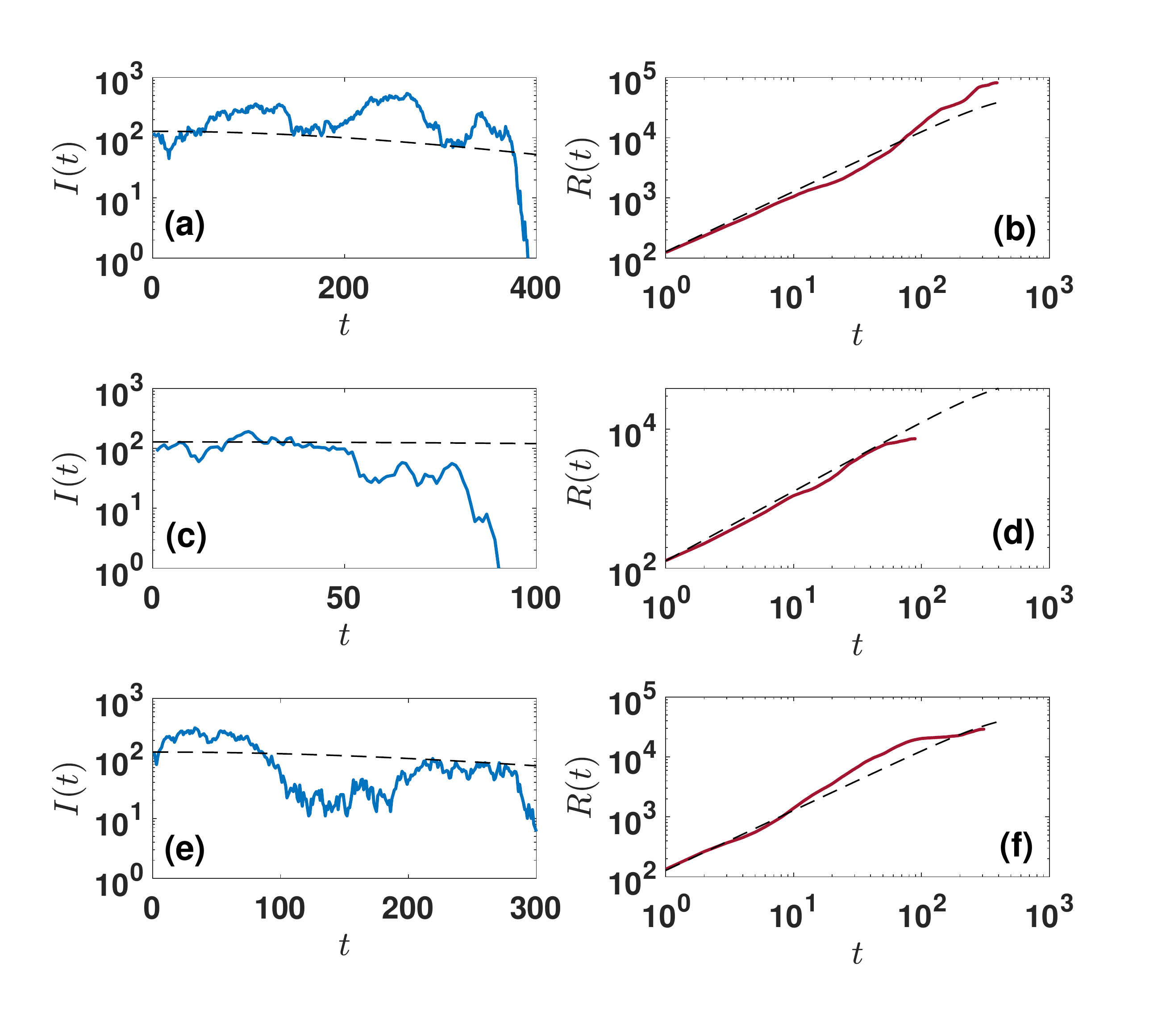}
        \end{center}
         \caption{We show three examples of  time series for the number
           of infected individuals $I(t)$ (panels a, c and e) and the corresponding  number of removed
           individuals $R(t)$ (panels b, d and f) for a critical SIR dynamics with
           non-trivial initial condition. The time series are obtained
           by simulating the  stochastic SIR dynamics at criticality (with $\lambda=1$) on a
           well-mixed population with identical parameters: initial
           number of infected individuals $n_0=128$ and population
           size $N=10^7$. The dashed lines indicate the corresponding
           deterministic predictions. }
  	\label{fig:time series}
      \end{figure}
      
      Please note that all the above questions cannot be answered with
      a purely deterministic approach. SIR outbreak sizes and
      durations obey probability distributions that are well peaked
      around their expected value only if the system is off
      criticality. However, the very fact that the system
      is assumed to be in the critical regime implies that
      fluctuations have a dominant role in the determination of the
      properties of the dynamical system.
      In Figure $\ref{fig:time series}$ for example, we display time
      series representative for the critical regime of the dynamics.
      Ground-truth time series are obtained by simulating the SIR
      stochastic dynamics (see Appendix A for details).
      They are compared with the deterministic expectation value obtained
      by integrating Eqs.~(\ref{sys}). We note that some realizations
      of the process are more persistent and more pervasive in the population than what predicted by the expected value.

From here on, we abandon the deterministic SIR equations and we embrace
a stochastic approach. Critical SIR dynamics starting from a single
initial seed, i.e., $n_0=1$, is known to be characterized by extremely large
fluctuations of the outbreak size and duration. These fluctuations can
be quantified by leveraging the mapping between critical SIR in a
well-mixed population and the mean-field branching process.  In the
following sections, we first review results valid for
$n_0=1$. Then, we focus our attention on the non-trivial case $n_0> 1$.

\subsection{Critical dynamics with  $n_0=1$ initial seed }

If the initial condition is such that only one node is  in the infected state while all other nodes are in the susceptible state, the critical SIR model gives rise to outbreaks that follow the statistics of a critical branching process~\cite{zapperi1,zapperi2} corrected by some scaling functions $F_T(N/N_T^{\star})$ and $F_R(N/N_R^{\star})$ that implement the
effective cutoff caused by finite-size effects~\cite{uno,due}. Here, $N$ is the size of the system; $N_T^{\star}$ and $N_R^{\star}$ are instead parameters that determine when the cutoff takes place.
Specifically,   the distribution $P(T)$ of the duration $T$ of an outbreak follows the law
\bea
P(T)\sim T^{-2}F_T(N/N_T^{\star}),
\label{Pt}
\eea
while the size of the outbreak $R$ follows the distribution 
\bea
P(R)\sim R^{-3/2}F_R(N/N_R^{\star}).
\label{Pr}
\eea

The cutoff sizes $N_T^{\star}$ and $N_R^{\star}$ have been derived in Refs.~\cite{uno,due}. They are given by 
\bea
N_T^{\star}=T^{3}, & N_R^{\star}=R^{3/2}.
\eea
From the expressions for  $P(T)$ and $P(R)$ given by Eqs.~(\ref{Pt}) and~(\ref{Pr}), respectively, and further assuming a sharp cutoff, it is easy to deduce that the scaling with the system size $N$ of the average outbreak size $\avg{R}$, the average duration $\avg{T}$, and the standard deviations $\sigma_R$ and $\sigma_T$~\cite{uno,due} obey
\bea
\avg{R}\sim N^{1/3}, &
\sigma_R\sim N^{1/2},\nonumber \\
\avg{T}\sim \ln N, & \sigma_T\sim N^{1/6}.
\eea
We observe that all the above quantities are sub-extensive, as they
all grow sub-linearly with the system size.
The expected critical outbreak size $\avg{R}$ grows as the system
size to the power of $1/3$. However, the standard deviation associated
to the outbreak size, i.e., $\sigma_R$, grows with increasing system size much faster than $\avg{R}$.
This fact indicates that it is very challenging to make predictions if the dynamics is critical.
Similarly, the outbreak duration is characterized by large fluctuations  in the large population limit.
We note that the exponents $2$ and $3/2$ of the distribution $P(T)$
and $P(R)$ are the critical mean-field exponents. These
exponents are universal and are observed for many critical spreading
processes~\cite{radicchi2020classes}.
They characterize the critical SIR
on network topologies too
as long as the underlying network has a homogeneous degree
distribution. In power-law networks,
these exponents can deviate from their mean-field values
as investigated in Refs.~\cite{Goh, radicchi2020classes}.

We have seen in Section II that the deterministic approach predicts a
linear increase of the number of removed individuals
with time for small time. However, such a prediction is not accurate for the
ground-truth dynamics; accounting for stochastic effects correctly
predicts a quadratic growth
of the number of removed individuals in time when the epidemic starts
with a single initial seed.
To this end, the
number of removed individuals grows in time as a power law
\bea
R=\overline{t}^z,
\label{dz}
\eea
where $z$ is a dynamical critical exponent, and $\overline{t}$ is the
expectation value of the time necessary to observe $R$ removed individuals.
The value of the dynamical critical exponent can be obtained in
different ways~\cite{zapperi2}. Here, we present the derivation of the
value of the dynamical exponent based on Langevin-like equations for
the dynamics. Starting from an initial fraction of infected individuals
$ i(0)=n_0/N$ and a fraction $r(0)=0$ of removed individuals we write
\bea
\frac{di}{dt}&=&(\lambda s-1) i+c\sqrt{i}\eta(t),\nonumber \\
\frac{dr}{dt}&=&i,
\eea
where $\eta(t)$ is an uncorrelated white noise with
$\mathbb{E}({\eta(t)})=0$
and $\mathbb{E}(\eta(t)\eta(t'))=\delta(t-t')$ and $c$ is a constant.
At criticality, i.e., $\lambda=\lambda_c=1$, thus, assuming $t\ll1$
and $i(0)\ll 1$, we have $\lambda s-1\simeq -i(0)$.  We can therefore write 
\bea
\frac{di}{dt}&=&c\sqrt{i}\eta(t),\nonumber \\
\frac{dr}{dt}&=&i.
\label{stoc}
\eea
We now perform a simple scaling analysis of this stochastic equations
as usually done in non-equilibrium statistical mechanics, e.g., Refs.~\cite{barabasi1995fractal,Hinrichsen,Marro}.
If we rescale time as
\bea
t\to bt
\eea
and define the scaling exponents $z,\alpha,$ for $r,i$
\bea
r & \to & b^z t,\nonumber \\
i & \to &b^{\alpha} t,\\
\eea
as the exponents that leave the SIR critical dynamics unchanged.
The SIR stochastic Eqs.~(\ref{stoc}) read 
\bea
b^{\alpha-1}\frac{di}{dt}&=&c b^{\alpha/2-1/2}\eta(t),\nonumber \\
b^{z-1}\frac{dr}{dt}&=&b^{\alpha}i ,
\eea
from which we can derive the scaling exponents 
\bea
\alpha &=&1\\
z&=&2.
\eea

In summary, in the critical dynamical regime, if the spreading is
started from a single initial seed, we expect that the average number
of removed individuals grows quadratically with time.

\subsection{Critical dynamics with $n_0>1$ initial seeds}

\begin{figure}[h!]
	\begin{center}
 \includegraphics[width=0.98\columnwidth]{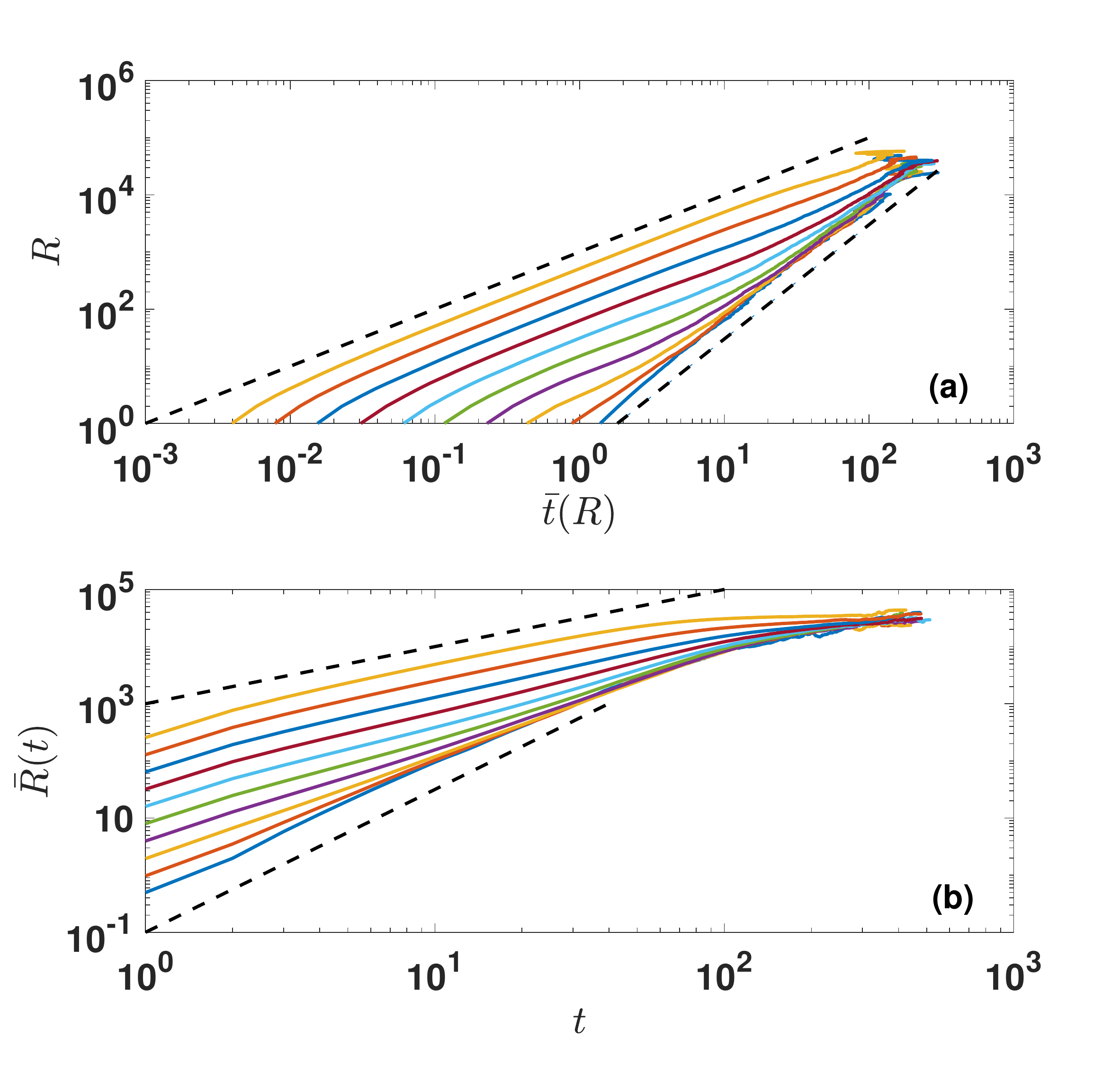}	
  \end{center}\caption{Dynamical properties of the critical SIR
    process with non-trivial initial condition.
    a) We plot the number of removed individuals $R$ versus
    the expected time $\bar{t}(R)$ required to observe $R$ removed
    individuals. The different curves indicate different initial
    conditions, from bottom to top,
    $n_0=2^q$,  with $q = 2, 3, 4, \ldots, 8, 9$.
    The population size is
    $N=10^7$. The dashed lines are guides to the eye and correspond to
    linear and quadratic growth of $R$ versus $\bar{t}(R)$, respectively. b) Expected number $\bar{R}(t)$ of removed individuals
    as a function of time $t$. Data are the same as in panel a. The
    dashed lines are guides to the eye and correspond to power-law
    growth of $\bar{R}$ versus $t$ with exponent $\xi=1$ and
    $\xi=2.5$, respectively.}
  	\label{fig:dynamics}
	\end{figure}

Critical SIR dynamics started from the non-trivial initial condition
$n_0>1$ differs from the
critical SIR dynamics started from $n_0=1$ seed.
To include an explicit dependence on
the parameter $n_0$ in the scaling of Eq.~(\ref{dz}), we 
correct it by introducing the scaling function $F(x)$, where
$x=n_0/\overline{t}(R)$. We impose that 
\bea
R\simeq \overline{t}(R)^zF\left(\frac{n_0}{\overline{t}(R)}\right)
\eea
with 
\bea
F(u)\sim\left\{\begin{array}{ccc}1& \mbox{if} &u\gg 1 \\
u^{\beta} & \mbox{if}& u\ll 1 \end{array}\right.
\eea
According to the deterministic SIR equations for $t\ll1$ and
$i(0)\ll1$, the number of removed individuals
grows linearly in time with a slope $i(0)=n_0/N$.
Thus, we deduce that 
\bea
 \beta=1.
\eea
This value is well supported by extensive numerical results (see
Figure $\ref{fig:dynamics}$) which confirm that there is a cross-over between linear and quadratic dependence of $R$ on $\overline{t}(R)$.

In the simulation of the SIR model, it is natural to study the behavior
of $R$ as a function of $\overline{t}(R)$. However, in real epidemic
time series, the number of infected individuals is measured over
constant time intervals.
The two ways of monitoring the evolution of the process,  i.e.,  $R$ versus $\overline{t}(R)$ rather than $\overline{R}(t)$
versus time $t$,  may lead to the observation of
different scaling exponents. The discrepancy is due to the stochastic
nature of the spreading process. The phenomenon is apparent from the
results of Figure $\ref{fig:dynamics}$: depending on the type of
measurement performed on the system, the power-law increase of the number of removed individuals as a
function of time can be described by a continuous range of exponents
ranging from $\xi=1$ to $\xi\sim 2.5$.
We can therefore write
\bea
\overline{R}(t)\simeq n_0t^{\xi}h(t,N)
\eea
where $\xi$ is a decreasing function of $n_0$, and $h(t,N)$ is a
modulating function expressing the deviation from the pure power-law
behavior. The ansatz of the above equation is compatible with the power-law scaling
of the empirical time series of removed individuals as a function of time observed in  countries where containment measures have been implemented extensively~\cite{Ziff,Nekovee,Blasius,quadratic}.

 \subsection{Distribution of avalanche durations and  sizes for  the critical SIR model initiated by $n_0 > 1$ seeds}

\begin{figure}[h!]
	\begin{center}
 \includegraphics[width=0.98\columnwidth]{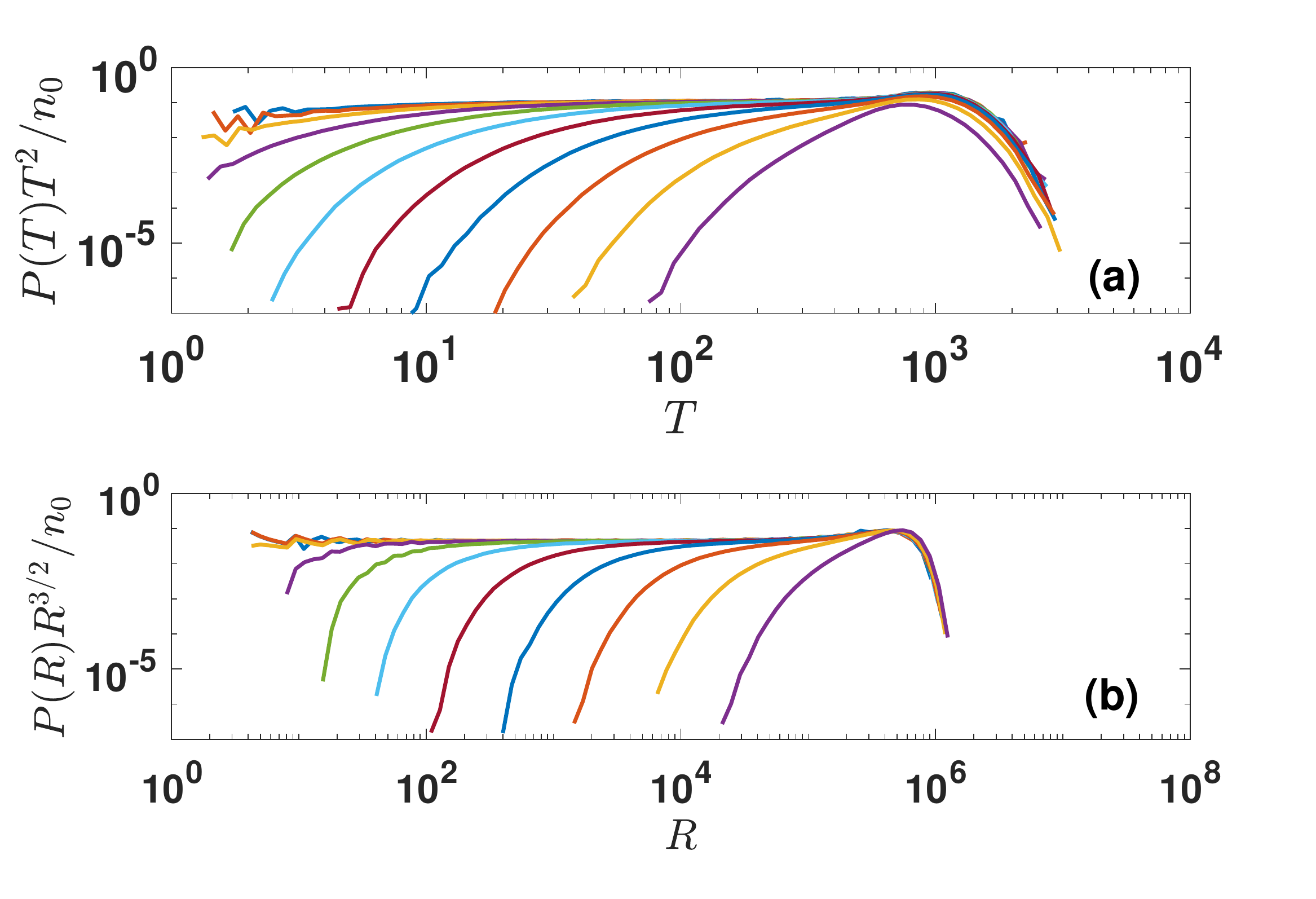}	
  \end{center}\caption{The rescaled distribution of outbreak duration
    $P(T)$ (panel a) and outbreak size $P(R)$ (panel b) are plotted
    for a well-mixed population of $N=10^8$ individuals
    and initial number of infected individuals $n_0$ equal to $2^q$,
 with $q = 2, 3, 4, \ldots, 10, 11$.
      The distributions display a lower cutoff that increases as $n_0$
      increases, and an upper cutoff whose value does not strongly
      depend on $n_0$. }
  	\label{fig:distribution}
	\end{figure}

\begin{figure}[h!]
	\begin{center}
 \includegraphics[width=0.98\columnwidth]{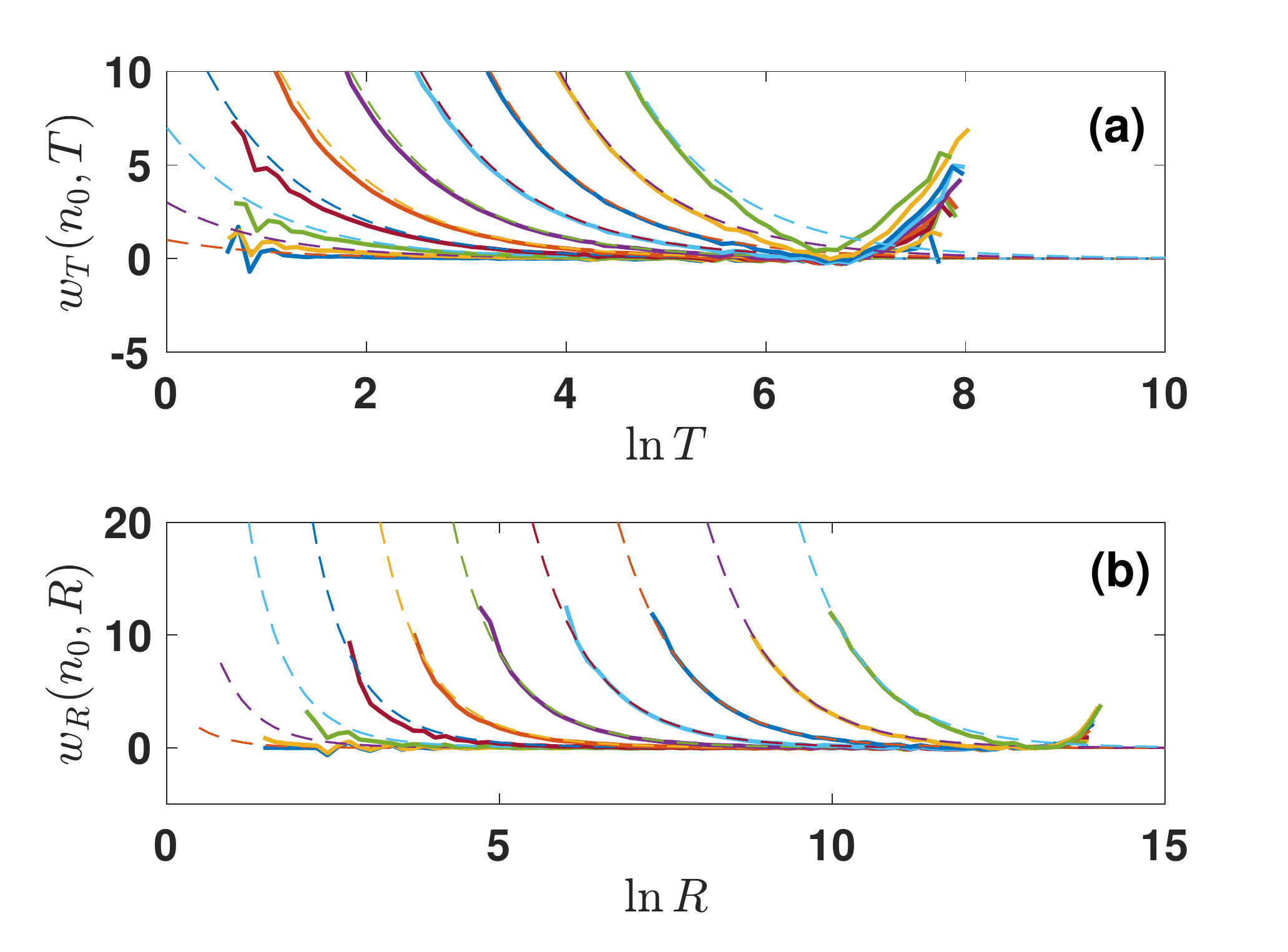}	
  \end{center}\caption{ The function $w_T(n_0,T)$  and $w_R(n_0,R)$
    defined in Eq. (\ref{wT}) (panel a) and Eq. (\ref{wR}) (panel b) are plotted for a
    population of $N=10^8$ individuals and different values of
    $n_0=2^q$,
with $q = 2, 3, 4, \ldots, 10, 11$.
    The dashed lines correspond to the scaling
    expressed  in  Eq. (\ref{WTt}) and
    Eq. (\ref{wRt}). 
}
  	\label{fig:w}
	\end{figure}

 In this section we investigate the statistical properties of the
 distribution of outbreak duration and size for the critical  SIR
 dynamics in a well mixed-population when the initial condition is non-trivial, i.e., $n_0>1$. 
 Scaling arguments  suggest the following expression for  the distribution $P(T)$ of the critical outbreak duration $T$ 
\bea
P(T)\sim n_0 T^{-2} {F}_T(N/N_T^{\star},n_0/T).
\label{Pt2}
\eea
The above scaling function is a natural modification of
Eq.~(\ref{Pt}) by assuming that $n_0$ scales like time.
In particular, the distribution $P(T)$ is characterized by a lower
cutoff depending on $n_0$. This fact is intuitive as an outbreak
with a larger number of initially infected individuals is not expected
to reach the absorbing state  faster than an
outbreak started by a single seed (see Figure
\ref{fig:distribution}a).
We note that, in the  critical SIR dynamics,
the dependence on $n_0$ does not lead to a change of the critical
exponent values, as for example observed in other non-equilibrium
phase
transitions~\cite{slip_exponent,Hinrichsen,hinrichsen1998correlated}. 
In Figure $\ref{fig:w}a$, we display the function 
\bea
w_T(n_0,T)=-\ln \frac{{F}_T(N/N_T^{\star},n_0/T)}{{F}_T(N/N_T^{\star},1/T)}
\label{wT}
\eea
and we demonstrate that the scaling function $w_T(n_0/T)$  for $n_0\ll N_T^{\star}$ can be approximated as
\bea
{w}_T(n_0,T)\simeq -\frac{n_0-1}{T}.
\label{WTt}
\eea
The scaling behavior, valid for $n_0 \ll N^{1/3}$, can be justified 
by  assuming that each of the $n_0$ seeds generates an independent
outbreak obeying the statistics of the critical branching process.
A critical avalanche started from a single infected individual has a
duration $T$ following the power-law distribution $\pi(T)\sim
T^{-2}$~\cite{zapperi1,zapperi2}. Thus, assuming independence among
the $n_0$ avalanches, we can estimate the probability $P(T)$ as the
probability that among all $n_0$ outbreaks   the last outbreak to get extinguished is extinguished at time $T$. Therefore in the infinite population limit we obtain
\bea
P(T)\simeq n_0 [1-\pi_c(T)]^{n_0-1} \pi(T),
\eea
where $\pi_c(T)$ is the probability that an outbreak generated by a single
infected individual 
is not extinguished at time $T$, with $\pi_c(T)\int_T^{\infty}\pi(x) dx\simeq{1}/{T}$.
By assuming $1\ll n_0\ll N^{1/3}$, we get
\bea
P(T)\simeq n_0\exp\left(-\frac{n_0-1}{T}\right).
\eea
Finally, we note that while the scaling behavior described in Eq. (\ref{WTt}) has strong numerical confirmation for $T\ll N_T^{\star}$ for values of  $T\sim N_T^{\star}$ the scaling function $w_T(n_0,T)$ signals a dependence of the cutoff on $n_0$ (see Figure $\ref{fig:w}$).

Scaling arguments suggest that the  distribution $P(R)$ of critical
outbreak size $R$ should obey 
\bea
P(R)\sim n_0 R^{-3/2}F_R(N/N_R^{\star},n_0,R) ,
\eea
where the function $F_R(N/N_R^{\star},n_0,R)$ implements a lower
cutoff dependent exponentially on $n_0$ (see Figure \ref{fig:distribution}b).

In Figure $\ref{fig:w}b$, we show the function 
\bea
w_R(n_0,R)=-\ln \frac{{F}_R(N/N_R^{\star},n_0,R)}{{F}_R(N/N_R^{\star},1,R)}
\label{wR}
\eea
which,  for $n_0\ll N_R^{\star}(R,n_0)$, can be approximated as
\bea
{w}_R(n_0,R)\simeq-\frac{n_0^2}{R}-\frac{n_0^{5/2}}{R^2} .
\label{wRt}
\eea
This scaling function
indicates that, for large values of $R$,
$R$ scales like $n_0^{2}$. For small values of $R$, it is possible to
observe some corrections would be required to fully describe the scaling.
We notice that, in the first order in $n_0$, the normalization constant of the distribution $P(R)$ is independent of $n_0$.
A way to interpret the result is by considering the infinite population limit approximating the distribution $P(R)$
as the convolution of the $n_0$ sizes of independent outbreak
events. In this limit  we have 
\bea
P(R)=\int d\omega e^{i\omega R} \left[\mathcal{F}(\omega)\right]^{n_0},
\label{ind}
\eea 
where $\mathcal{F}(\omega) =\sum_{r}\Pi(r)e^{-i\omega r}$ is the
generating function of the distribution $\Pi(r)$ of avalanches sizes
of SIR critical dynamics starting from a single seed. 
Assuming in first approximation that $\Pi(r)$ is a pure power law
$\Pi(r)\sim  r^{-3/2}$, it follows that the logarithm of the generating
function behaves, for small $\omega$,
as $\ln \mathcal{F}(\omega)\sim \sqrt{\omega}$. The result, together
with Eq.~(\ref{ind}), indicates  that $R$ should scale as $n_0^2$  for $n_0\ll N^{2/3}$. 
For mor details on the infinite population limit we refer the reader to Ref. \cite{krapivsky2020infection}.

 \subsection{Statistical properties of the critical outbreak started by $n_0 > 1$ seeds}

\subsubsection{General scenario}
\begin{figure}[h!]
		\begin{center}
 \includegraphics[width=0.98\columnwidth]{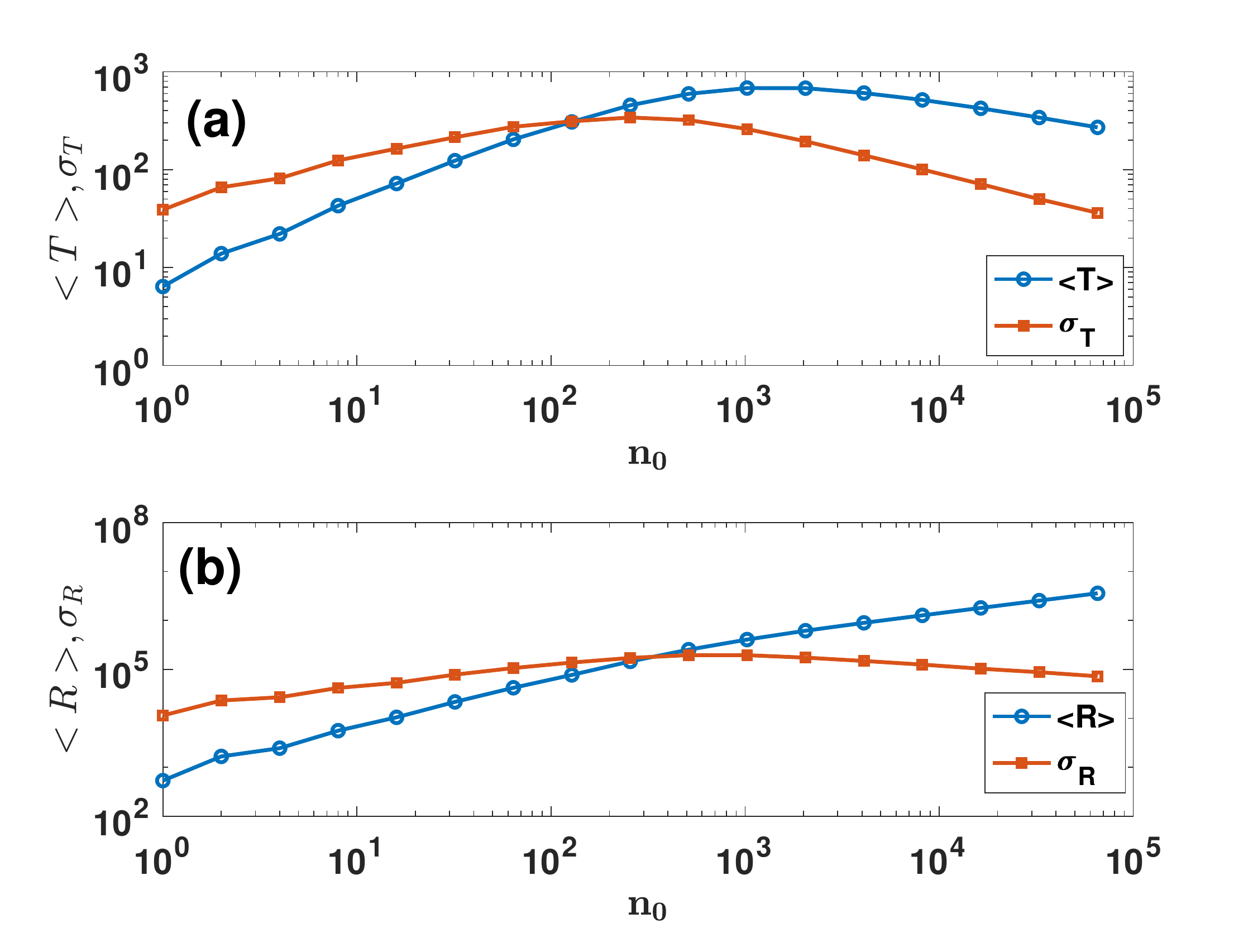}	
  \end{center}\caption{a) The expected value of the duration of
    critical outbreak $\avg{T}$ and the corresponding standard
    deviation $\sigma_T$ are plotted versus the number $n_0$ of
    individuals initially infected. The population size is $N =
    10^8$. Results are obtained relying on $10^4$ independent
    realizations of the process. b) The expected size of  the
    critical outbreak $\avg{R}$ and the corresponding standard
    deviation $\sigma_R$ are plotted versus the number $n_0$ of
    individuals initially infected. Data are the same as of panel a.}
  	\label{fig:stat}
	\end{figure}

 We performed large-scale simulations of the critical SIR model to
 address fundamental questions regarding the distributions of 
 duration $T$ and size $R$ of outbreaks started
 by a non-trivial initial condition $n_0>1$.  In
 Figure~$\ref{fig:stat}$, we display
 the average values and the standard deviations of both $T$ and $R$
 for a large system composed of $N=10^8$ individuals. We display the
 moments of the distributions as a function of the number of initial
 seeds $n_0$.  The main outcomes are as follows.
 The expected size $\avg{R}$ is a growing function of $n_0$.
 The expected duration $\avg{T}$ is a non-monotonic function of
 $n_0$, displaying a single peak. The standard deviations $\sigma_T$
 and $\sigma_R$ also display a peak as a function of $n_0$. The
 coefficient of variation
 $\sigma_T/\avg{T}$ and $\sigma_R/\avg{R}$ are monotonically decreasing with $n_0$.
We conclude that fluctuations are fundamental to properly characterize the critical dynamical
regime. This statement is true for any value of $n_0$, albeit, in
relative terms, the most severe effect of fluctuations is observed for $n_0=1$.
We note that as the initial number of infected individuals  $n_0$ increases, the expected size of the outbreak displays a monotonic increase while the expected duration of the outbreak displays a maximum.
In the following subsection, we will provide scaling laws for these major statistical properties of the critical dynamics as a function of the number of initially infected individuals.

	\subsubsection{Scaling analysis of $\avg{T}$ and $\sigma_T$}
\begin{figure}[h!]
	\begin{center}
 \includegraphics[width=0.99\columnwidth]{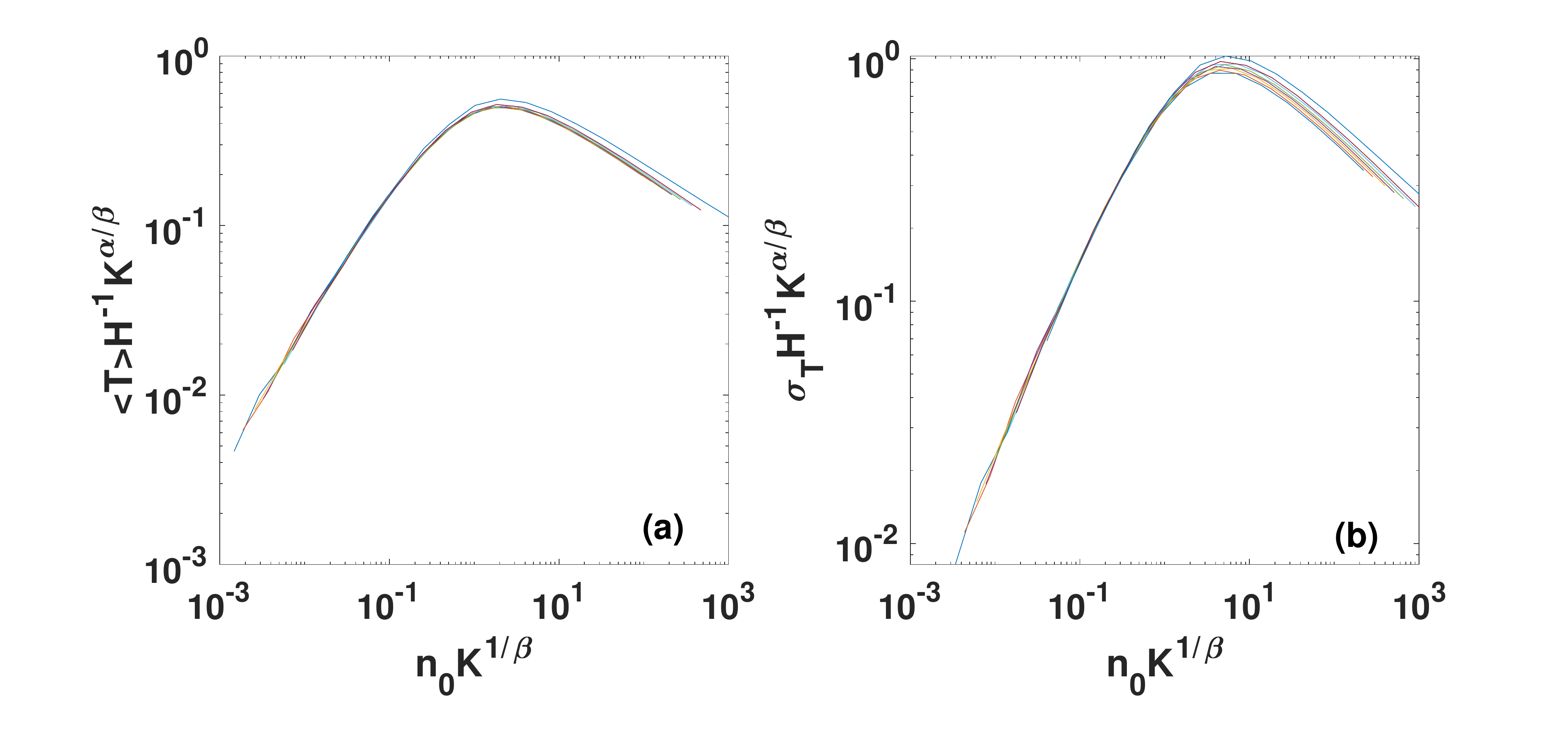}	
  \end{center}\caption{Finite-size scaling analysis for the duration
    $T$ of the critical SIR started from $n_0$ initial seeds. a) Data
    describing the average value of the outbreak duration  $\avg{T}$
    for different system sizes $N$ are collapsed on a unique universal
    curve using the scaling of Eq.~(\ref{collapseT}). Data are shown
    for population sizes $N$ ranging from $N=10^5$ to $10^8$. Each
    data point is obtained by simulating the SIR process $10^4$
    times. b) Same as in panel a, but for the standard deviation  $\sigma_T$.}
  	\label{fig:collapseT}
	\end{figure}

        We make the ansatz that the average duration $\avg{T}$ 
        can be described by
\bea
\avg{T}&\simeq & G(n_0|\alpha,\beta,H,K),
\label{scalingT}
\eea
where the function $G(n_0|\alpha,\beta,H,K)$ is given by 
\bea
G(n_0|\alpha,\beta,H,K)=\frac{n_0^{\alpha}H}{1+n_0^{\beta}K}.
\label{scalingT2}
\eea
Here, $\alpha$ and $\beta$ are, in the large population limit,
independent of $N$. On the contrary, $H$ and $K$ are dependent
on the population size.
By introducing the function 
\bea
g(x|\alpha,\beta)=\frac{x^{\alpha}}{1+x^{\beta}},
\label{gf}
\eea
we observe that it is possible to rescale the curves obtained for different values of $N$ by performing the transformation
\bea
H^{-1}K^{\alpha/\beta}G(n_0|\alpha,\beta,H,K)=g(n_0K^{1/\beta}|\alpha,\beta).
\label{collapseT}
\eea
This expression allows us to perform a data collapse of the
data obtained for $\avg{T}$ at different values of $n_0$
and different population size $N$ (see Figure~\ref{fig:collapseT}a).

We observe that, if we start from a non-trivial initial condition
$n_0$, the expected duration of the outbreak $\avg{T}$
reaches its maximum at 
\bea
n_0^{\star}=\left(\frac{\alpha}{K(\beta-\alpha)}\right)^{1/\beta}.
\label{nostar}
\eea 

\begin{figure}
	\begin{center}
 \includegraphics[width=0.9\columnwidth]{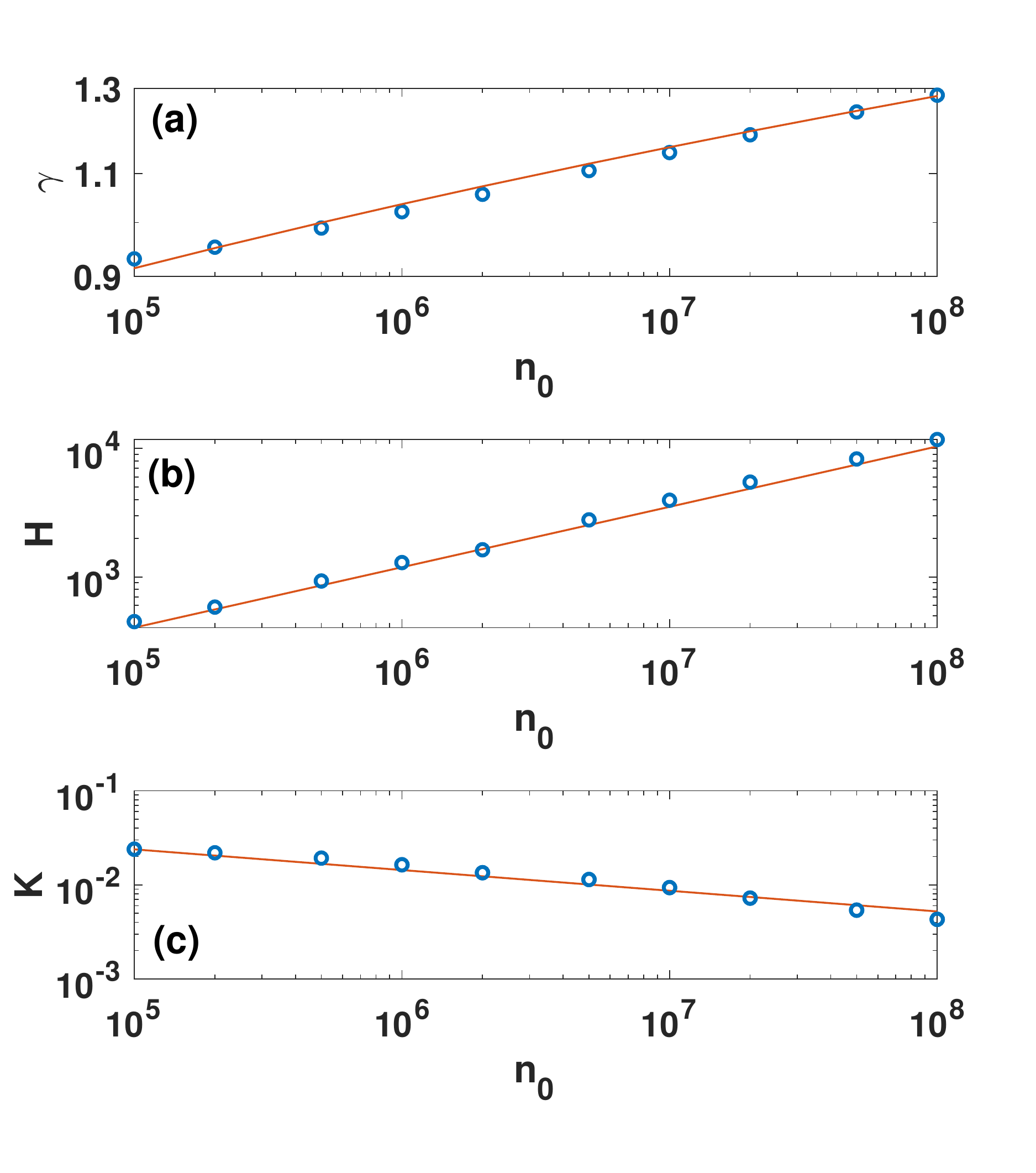}	
  \end{center}\caption{  a) The fitting parameter
    $\gamma$, i.e., Eq.(\ref{scalingR}), as a function of the system
    size $N$.  Data points are the same as those of
  Figure~\ref{fig:collapseT}. The line displays the best fit of the
  data points with Eq.~(\ref{gammaR}).  b) Scaling parameter $H$ as a function of the
  system size $N$ for the standard deviation $\sigma_{R}$ of critical
  outbreaks. The solid line corresponds to the scaling function of
  Eq.(\ref{sigmaR}). c) Same as in panel b, but for the
  scaling parameter $K$. The solid line corresponds to the scaling function of
  Eq.(\ref{sigmaR}).}
  	\label{fig:parametersR}
	\end{figure}

The scaling parameters  $H$ and $K$ obey the scaling relation
\bea
K\simeq a_K N^{\delta_K} ,& H\simeq a_H \log(N)+b_H.
\label{Tm}
\eea
with $a_H=0.58\pm 0.08$, $b_H=-2.0\pm 0.9$, $\delta_K=-0.37\pm 0.04$ and $a_K=0.75\pm 0.06$ (see Figure~\ref{fig:parametersR}).
We note that the logarithmic scaling of $H$ is expected from the known scaling of $\avg{T}$ for the SIR critical model starting from a single initial seed. 
The exponents $\alpha$ and $\beta$ are given by 
\bea
\alpha=0.78\pm 0.03, &\beta=1.10\pm 0.1 .
\eea

The standard deviation of the outbreak duration $\sigma_T$ can be described in the same exact way as $\avg{T}$. The
ansatz 
\bea
\sigma_{T}&\simeq & G(n_0|\alpha,\beta,H,K),
\label{scalingT3}
\eea
leads to the data collapse  shown in
Figure~\ref{fig:collapseT}b. The scaling parameters $H$ and $K$ obey the scaling relations
\bea
K\simeq a_K N^{\delta_K}, & H\simeq a_H N^{\delta_H}.
\label{sigmaT}
\eea
with $\delta_H=0.17\pm 0.01$, $b_H=1.7\pm 0.2$, $\delta_K=-0.36\pm
0.03$ and $a_K=2.3\pm 0.7$ (see Figure~\ref{fig:parametersR}).
We note that $\delta_H\simeq 1/6$. Therefore, for $n_0=1$ the scaling
reduces
to the well-known scaling for the critical SIR model starting from a single initial seed. 
Moreover, the exponent $\alpha$ and $\beta$ for $\sigma_T$ are given by 
\bea
\alpha=0.50\pm 0.05, &\beta=1.0\pm 0.1.
\eea

\begin{figure}[h!]
	\begin{center}
 \includegraphics[width=0.99\columnwidth]{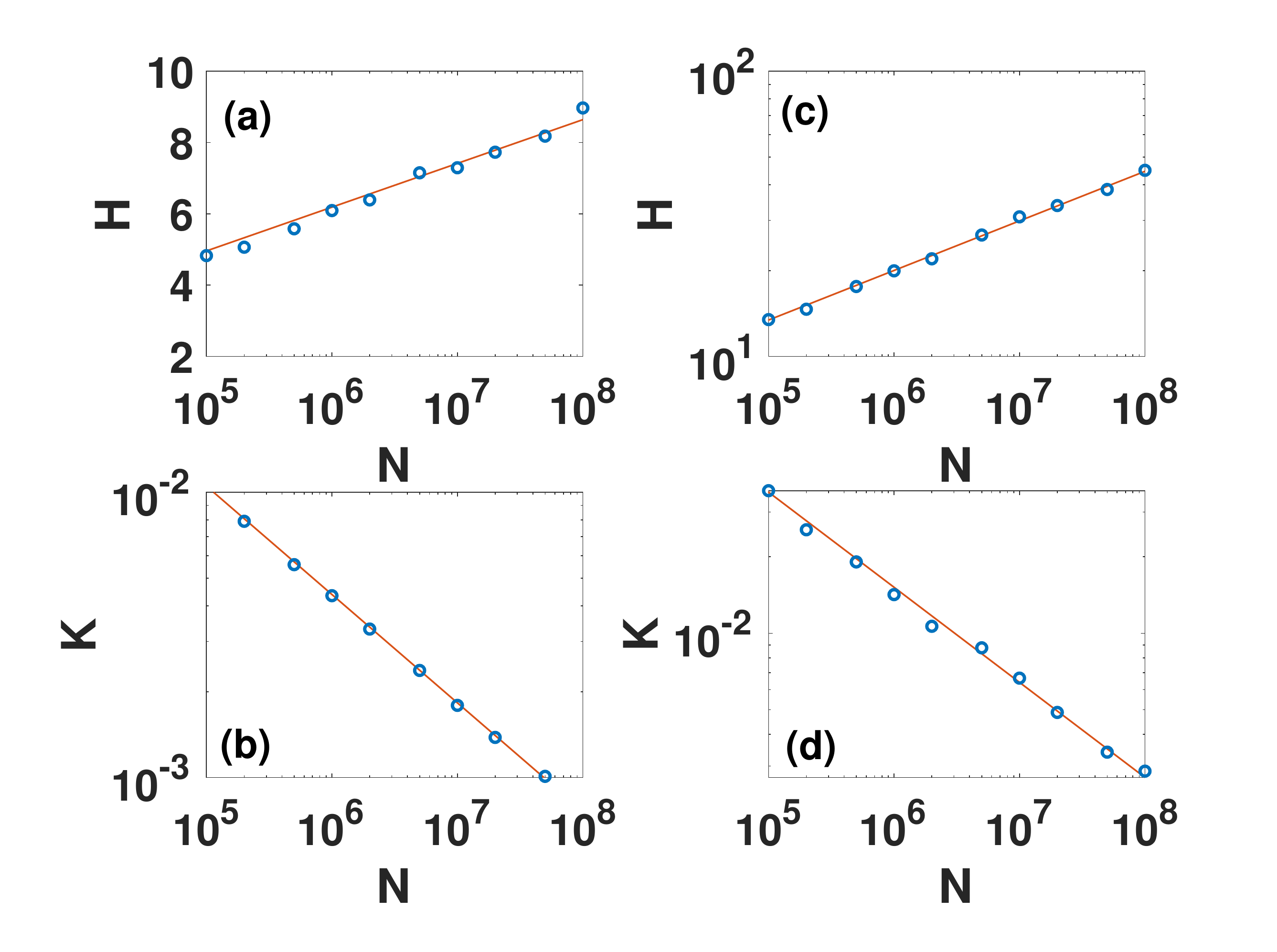}	
\end{center}\caption{a) Scaling parameter $H$ as a function of the
  system size $N$ for the average duration $\avg{T}$ of critical
  outbreaks. Data points are the same as those of
  Figure~\ref{fig:collapseT}. The line displays the best fit of the
  data points with Eq.~(\ref{Tm}). b) Same as in panel a, but for the
  scaling parameter $K$. c) Same as in panel a, but for the standard
  deviation $\sigma_T$. The orange line is the best fit of the data
  points with Eq.~(\ref{sigmaT}). d) Same as in panel c, but for the
  scaling parameter $K$.}
  	\label{fig:parametersT}
	\end{figure}

 \subsubsection{Scaling analysis of $\avg{R}$ and $\sigma_R$}
 
\begin{figure}
	\begin{center}
 \includegraphics[width=0.9\columnwidth]{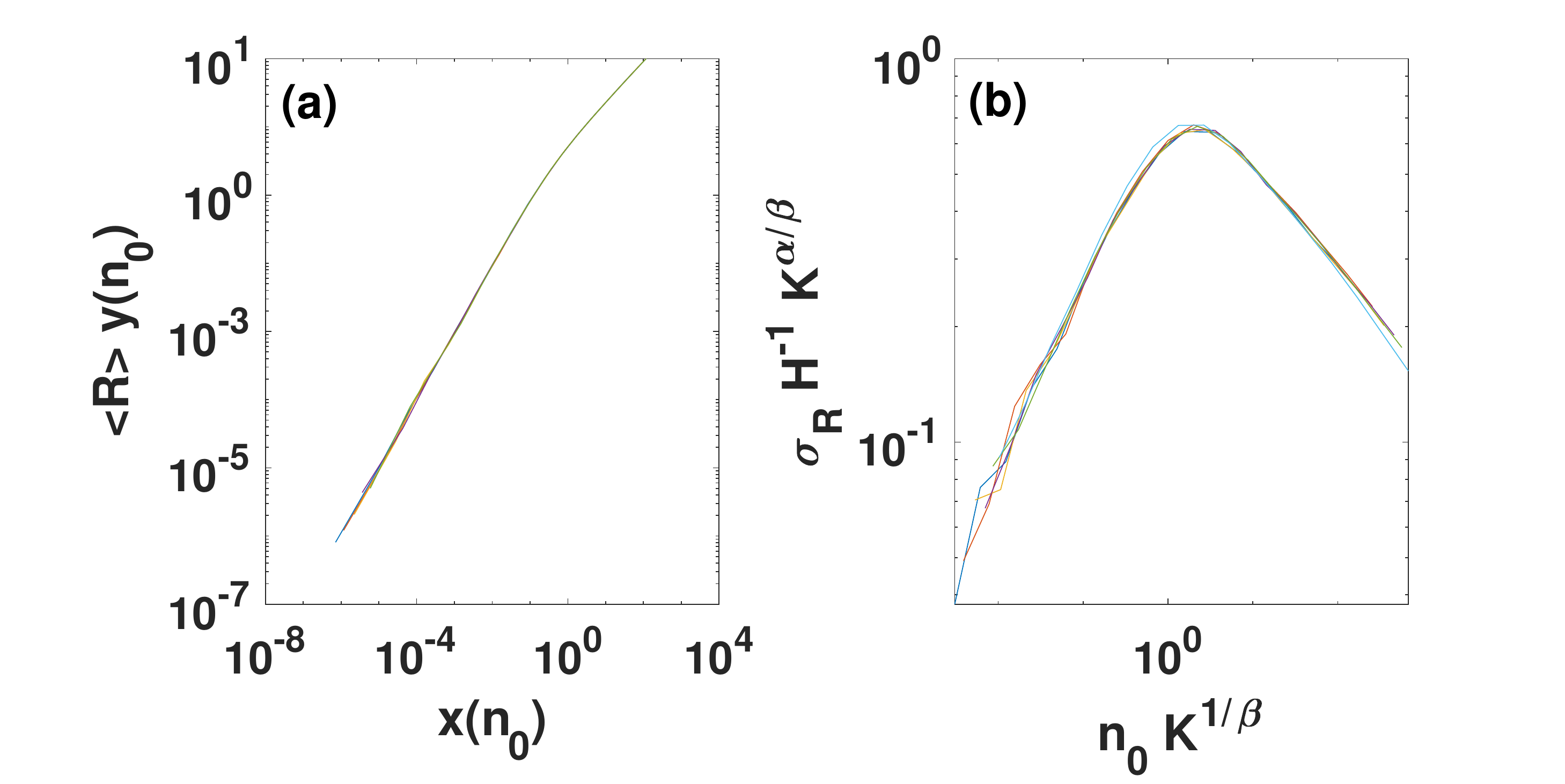}	
\end{center}\caption{
Finite-size scaling analysis for the size
    $R$ of the critical SIR started from $n_0$ initial seeds. a) Data
    describing the average value of the outbreak size  $\avg{R}$
    for different system sizes $N$ are collapsed on a unique universal
    curve using the scaling of Eq.~(\ref{collapseR}). Data are the
    same as of Figure~\ref{fig:collapseT}. b) Same as in panel a, but for the standard deviation  $\sigma_R$.}
  	\label{fig:collapseR}
      \end{figure}

The ansatz for $\avg{R}$ is slightly different from the one appearing
in Eq.~(\ref{scalingT2}), as it includes an additional logarithmic correction 
\bea
\avg{R}& \simeq & \frac{n_0^{\alpha}H}{1+n_0^{\beta}(\ln n_0)^{\gamma}K}.
\label{scalingR}
\eea
We  take  \bea
H=\frac{3}{2}N^{1/3},& K=N^{-1/3}
\eea
and 
\bea
\alpha=1, &\beta=0.5.
\eea
and we perform a fit of the exponent $\gamma$.
As   Figure~\ref{fig:parametersR}a and Figure~\ref{fig:collapseR}a  demonstrate,
the function gives rise to  excellent data fits as long as  the exponent $\gamma$ is
\bea
\gamma=a_\gamma\ln N+b_{\gamma}
\label{gammaR}
\eea
with $a_{\gamma}=0.053\pm 0.003$ and $b_{\gamma}=0.30\pm 0.06$.

 The function $\avg{R}$ can be rescaled and the data obtained for
 different $N$ collapsed on a universal curve (see
 Figure~\ref{fig:collapseR}). This task is
 done by noting that 
 \bea
 \avg{R}y(n_0)=g(x(n_0)|\alpha,\beta),
 \label{collapseR}
 \eea
 where \bea
 x(n_0)&=&n_0(\ln n_0)^{\gamma/\beta}K^{1/\beta},\nonumber \\
y(n_0)&=&H^{-1}K^{\alpha/\beta}(\ln n_0)^{\alpha \gamma/\beta},
 \eea
  and the function $g(x|\alpha,\beta)$ is given by Eq.~(\ref{gf}).
 The expected size of the outbreak $\avg{R}$ does not display a maximum as a function of $n_0$, i.e., it is a monotonous increasing function of $n_0$.
 The standard deviation $\sigma_R$ can be instead fitted using the same ansatz as $\sigma_T$, 
 i.e.,
\bea
\sigma_R&\simeq &G(n_0|\alpha,\beta,H,K).
\label{scalingR2}
\eea
  The best estimates of the parameters
are
\bea
\alpha=0.57\pm0.03, &\beta=0.95\pm 0.05,
\eea
 and 
\bea
K\simeq a_K N^{\delta_K}, & H\simeq a_H N^{\delta_H}.
\label{sigmaR}
\eea
with $\delta_H=0.48\pm 0.02$, $b_H=1.8\pm 0.5$, $\delta_K=-0.25\pm
0.03$ and $a_K=0.49\pm 0.3$
(see Figure~\ref{fig:parametersR}). The corresponding data collapse is shown in Figure~\ref{fig:collapseR}b. 

\section{Conclusions}
Motivated by the current COVID-19 pandemic, we have investigated the
critical properties of the Susceptible-Infected-Removed (SIR)
dynamics in well-mixed populations starting from
non-trivial initial conditions consisting of $n_0>1$ infected
individuals.
Although the modeling framework oversimplifies the real-world
scenario,  the setting is realistic in two main respects. First, the plateauing
time series observed in empirical data are compatible with the
critical dynamical regime. Second, the initial condition $n_0>1$ is
representative for a critical regime reached, thanks to effective containment
measures, after that a significant
community transmission already  took place.
We have shown that a non-trivial initial condition $n_0>1$ introduces another typical scale on the dynamics inducing a lower cutoff in the distributions of the duration and size of critical outbreaks.
The critical dynamics is characterized by very strong fluctuations,
but the presence of a non-trivial initial condition mitigates the role
of the fluctuations. In particular, while for a single initial seed
the standard deviation on the outbreak size and duration is much
larger than the corresponding expectation values, the relative error
diminishes as the size $n_0$ of the initial seed set increases.
Moreover, numerical results indicate that, as the initial number of infected individuals  $n_0$ increases, the expected size of the outbreak increases while the expected duration first increases and then decreases, displaying a maximum. Using scaling arguments, and extensive numerical simulations we have deduced the scaling of the maximum duration and the corresponding number of initially infected individuals.

\section*{Acknowledgements}
We thank P. L. Krapivsky, Geza Odor and R. M. Ziff for interesting
discussions.
F. R. acknowledges support from the National Science Foundation (CMMI-1552487).
\bibliographystyle{apsrev4-1}
\bibliography{references}

\begin{thebibliography}{46}%
\makeatletter
\providecommand \@ifxundefined [1]{%
 \@ifx{#1\undefined}
}%
\providecommand \@ifnum [1]{%
 \ifnum #1\expandafter \@firstoftwo
 \else \expandafter \@secondoftwo
 \fi
}%
\providecommand \@ifx [1]{%
 \ifx #1\expandafter \@firstoftwo
 \else \expandafter \@secondoftwo
 \fi
}%
\providecommand \natexlab [1]{#1}%
\providecommand \enquote  [1]{``#1''}%
\providecommand \bibnamefont  [1]{#1}%
\providecommand \bibfnamefont [1]{#1}%
\providecommand \citenamefont [1]{#1}%
\providecommand \href@noop [0]{\@secondoftwo}%
\providecommand \href [0]{\begingroup \@sanitize@url \@href}%
\providecommand \@href[1]{\@@startlink{#1}\@@href}%
\providecommand \@@href[1]{\endgroup#1\@@endlink}%
\providecommand \@sanitize@url [0]{\catcode `\\12\catcode `\$12\catcode
  `\&12\catcode `\#12\catcode `\^12\catcode `\_12\catcode `\%12\relax}%
\providecommand \@@startlink[1]{}%
\providecommand \@@endlink[0]{}%
\providecommand \url  [0]{\begingroup\@sanitize@url \@url }%
\providecommand \@url [1]{\endgroup\@href {#1}{\urlprefix }}%
\providecommand \urlprefix  [0]{URL }%
\providecommand \Eprint [0]{\href }%
\providecommand \doibase [0]{http://dx.doi.org/}%
\providecommand \selectlanguage [0]{\@gobble}%
\providecommand \bibinfo  [0]{\@secondoftwo}%
\providecommand \bibfield  [0]{\@secondoftwo}%
\providecommand \translation [1]{[#1]}%
\providecommand \BibitemOpen [0]{}%
\providecommand \bibitemStop [0]{}%
\providecommand \bibitemNoStop [0]{.\EOS\space}%
\providecommand \EOS [0]{\spacefactor3000\relax}%
\providecommand \BibitemShut  [1]{\csname bibitem#1\endcsname}%
\let\auto@bib@innerbib\@empty
\bibitem [{\citenamefont {Anderson}\ \emph {et~al.}(1992)\citenamefont
  {Anderson}, \citenamefont {Anderson},\ and\ \citenamefont
  {May}}]{anderson1992infectious}%
  \BibitemOpen
  \bibfield  {author} {\bibinfo {author} {\bibfnamefont {R.~M.}\ \bibnamefont
  {Anderson}}, \bibinfo {author} {\bibfnamefont {B.}~\bibnamefont {Anderson}},
  \ and\ \bibinfo {author} {\bibfnamefont {R.~M.}\ \bibnamefont {May}},\
  }\href@noop {} {\emph {\bibinfo {title} {Infectious diseases of humans:
  dynamics and control}}}\ (\bibinfo  {publisher} {Oxford university press},\
  \bibinfo {year} {1992})\BibitemShut {NoStop}%
\bibitem [{\citenamefont {Maslov}\ and\ \citenamefont
  {Goldenfeld}(2020)}]{Nigel}%
  \BibitemOpen
  \bibfield  {author} {\bibinfo {author} {\bibfnamefont {S.}~\bibnamefont
  {Maslov}}\ and\ \bibinfo {author} {\bibfnamefont {N.}~\bibnamefont
  {Goldenfeld}},\ }\href@noop {} {\bibfield  {journal} {\bibinfo  {journal}
  {arXiv:2003.09564}\ } (\bibinfo {year} {2020})}\BibitemShut {NoStop}%
\bibitem [{\citenamefont {Wong}\ \emph {et~al.}(2020)\citenamefont {Wong},
  \citenamefont {Weiner}, \citenamefont {Tkachenko}, \citenamefont {Elbanna},
  \citenamefont {Maslov},\ and\ \citenamefont {Goldenfeld}}]{Nigel2}%
  \BibitemOpen
  \bibfield  {author} {\bibinfo {author} {\bibfnamefont {G.~N.}\ \bibnamefont
  {Wong}}, \bibinfo {author} {\bibfnamefont {Z.~J.}\ \bibnamefont {Weiner}},
  \bibinfo {author} {\bibfnamefont {A.~V.}\ \bibnamefont {Tkachenko}}, \bibinfo
  {author} {\bibfnamefont {A.}~\bibnamefont {Elbanna}}, \bibinfo {author}
  {\bibfnamefont {S.}~\bibnamefont {Maslov}}, \ and\ \bibinfo {author}
  {\bibfnamefont {N.}~\bibnamefont {Goldenfeld}},\ }\href@noop {} {\bibfield
  {journal} {\bibinfo  {journal} {arXiv preprint arXiv:2006.02036}\ } (\bibinfo
  {year} {2020})}\BibitemShut {NoStop}%
\bibitem [{\citenamefont {Ferretti}\ \emph {et~al.}(2020)\citenamefont
  {Ferretti}, \citenamefont {Wymant}, \citenamefont {Kendall}, \citenamefont
  {Zhao}, \citenamefont {Nurtay}, \citenamefont {Bonsall},\ and\ \citenamefont
  {Fraser}}]{ferretti}%
  \BibitemOpen
  \bibfield  {author} {\bibinfo {author} {\bibfnamefont {L.}~\bibnamefont
  {Ferretti}}, \bibinfo {author} {\bibfnamefont {C.}~\bibnamefont {Wymant}},
  \bibinfo {author} {\bibfnamefont {M.}~\bibnamefont {Kendall}}, \bibinfo
  {author} {\bibfnamefont {L.}~\bibnamefont {Zhao}}, \bibinfo {author}
  {\bibfnamefont {A.}~\bibnamefont {Nurtay}}, \bibinfo {author} {\bibfnamefont
  {D.~G.}\ \bibnamefont {Bonsall}}, \ and\ \bibinfo {author} {\bibfnamefont
  {C.}~\bibnamefont {Fraser}},\ }\href {\doibase 10.1126/science.abb6936}
  {\bibfield  {journal} {\bibinfo  {journal} {Science}\ } (\bibinfo {year}
  {2020}),\ 10.1126/science.abb6936}\BibitemShut {NoStop}%
\bibitem [{\citenamefont {Bianconi}\ \emph
  {et~al.}(2020{\natexlab{a}})\citenamefont {Bianconi}, \citenamefont {Sun},
  \citenamefont {Rapisardi},\ and\ \citenamefont
  {Arenas}}]{bianconi2020message}%
  \BibitemOpen
  \bibfield  {author} {\bibinfo {author} {\bibfnamefont {G.}~\bibnamefont
  {Bianconi}}, \bibinfo {author} {\bibfnamefont {H.}~\bibnamefont {Sun}},
  \bibinfo {author} {\bibfnamefont {G.}~\bibnamefont {Rapisardi}}, \ and\
  \bibinfo {author} {\bibfnamefont {A.}~\bibnamefont {Arenas}},\ }\href@noop {}
  {\bibfield  {journal} {\bibinfo  {journal} {arXiv preprint arXiv:2007.05277}\
  } (\bibinfo {year} {2020}{\natexlab{a}})}\BibitemShut {NoStop}%
\bibitem [{\citenamefont {Fanelli}\ and\ \citenamefont
  {Piazza}(2020)}]{fanelli}%
  \BibitemOpen
  \bibfield  {author} {\bibinfo {author} {\bibfnamefont {D.}~\bibnamefont
  {Fanelli}}\ and\ \bibinfo {author} {\bibfnamefont {F.}~\bibnamefont
  {Piazza}},\ }\href@noop {} {\bibfield  {journal} {\bibinfo  {journal} {Chaos,
  Solitons \& Fractals}\ }\textbf {\bibinfo {volume} {134}},\ \bibinfo {pages}
  {109761} (\bibinfo {year} {2020})}\BibitemShut {NoStop}%
\bibitem [{\citenamefont {Carletti}\ \emph {et~al.}(2020)\citenamefont
  {Carletti}, \citenamefont {Fanelli},\ and\ \citenamefont {Piazza}}]{timoteo}%
  \BibitemOpen
  \bibfield  {author} {\bibinfo {author} {\bibfnamefont {T.}~\bibnamefont
  {Carletti}}, \bibinfo {author} {\bibfnamefont {D.}~\bibnamefont {Fanelli}}, \
  and\ \bibinfo {author} {\bibfnamefont {F.}~\bibnamefont {Piazza}},\
  }\href@noop {} {\bibfield  {journal} {\bibinfo  {journal} {arXiv preprint
  arXiv:2005.11085}\ } (\bibinfo {year} {2020})}\BibitemShut {NoStop}%
\bibitem [{\citenamefont {Bianconi}\ \emph
  {et~al.}(2020{\natexlab{b}})\citenamefont {Bianconi}, \citenamefont
  {Marcelli}, \citenamefont {Campi},\ and\ \citenamefont {Perali}}]{bianconi}%
  \BibitemOpen
  \bibfield  {author} {\bibinfo {author} {\bibfnamefont {A.}~\bibnamefont
  {Bianconi}}, \bibinfo {author} {\bibfnamefont {A.}~\bibnamefont {Marcelli}},
  \bibinfo {author} {\bibfnamefont {G.}~\bibnamefont {Campi}}, \ and\ \bibinfo
  {author} {\bibfnamefont {A.}~\bibnamefont {Perali}},\ }\href@noop {}
  {\bibfield  {journal} {\bibinfo  {journal} {arXiv preprint arXiv:2004.04604}\
  } (\bibinfo {year} {2020}{\natexlab{b}})}\BibitemShut {NoStop}%
\bibitem [{\citenamefont {Bradde}\ \emph {et~al.}(2020)\citenamefont {Bradde},
  \citenamefont {Cerruti},\ and\ \citenamefont {Bouchaud}}]{bradde}%
  \BibitemOpen
  \bibfield  {author} {\bibinfo {author} {\bibfnamefont {S.}~\bibnamefont
  {Bradde}}, \bibinfo {author} {\bibfnamefont {B.}~\bibnamefont {Cerruti}}, \
  and\ \bibinfo {author} {\bibfnamefont {J.-P.}\ \bibnamefont {Bouchaud}},\
  }\href@noop {} {\bibfield  {journal} {\bibinfo  {journal} {arXiv preprint
  arXiv:2006.09829}\ } (\bibinfo {year} {2020})}\BibitemShut {NoStop}%
\bibitem [{\citenamefont {Maheshwari}\ and\ \citenamefont
  {Albert}(2020)}]{Reka}%
  \BibitemOpen
  \bibfield  {author} {\bibinfo {author} {\bibfnamefont {P.}~\bibnamefont
  {Maheshwari}}\ and\ \bibinfo {author} {\bibfnamefont {R.}~\bibnamefont
  {Albert}},\ }\href@noop {} {\bibfield  {journal} {\bibinfo  {journal} {arXiv
  preprint arXiv:2006.09189}\ } (\bibinfo {year} {2020})}\BibitemShut {NoStop}%
\bibitem [{\citenamefont {Arenas}\ \emph {et~al.}(2020)\citenamefont {Arenas},
  \citenamefont {Cota}, \citenamefont {Gomez-Gardenes}, \citenamefont
  {G{\'o}mez}, \citenamefont {Granell}, \citenamefont {Matamalas},
  \citenamefont {Soriano-Panos},\ and\ \citenamefont {Steinegger}}]{Arenas}%
  \BibitemOpen
  \bibfield  {author} {\bibinfo {author} {\bibfnamefont {A.}~\bibnamefont
  {Arenas}}, \bibinfo {author} {\bibfnamefont {W.}~\bibnamefont {Cota}},
  \bibinfo {author} {\bibfnamefont {J.}~\bibnamefont {Gomez-Gardenes}},
  \bibinfo {author} {\bibfnamefont {S.}~\bibnamefont {G{\'o}mez}}, \bibinfo
  {author} {\bibfnamefont {C.}~\bibnamefont {Granell}}, \bibinfo {author}
  {\bibfnamefont {J.~T.}\ \bibnamefont {Matamalas}}, \bibinfo {author}
  {\bibfnamefont {D.}~\bibnamefont {Soriano-Panos}}, \ and\ \bibinfo {author}
  {\bibfnamefont {B.}~\bibnamefont {Steinegger}},\ }\href@noop {} {\bibfield
  {journal} {\bibinfo  {journal} {MedRxiv}\ } (\bibinfo {year}
  {2020})}\BibitemShut {NoStop}%
\bibitem [{\citenamefont {Ziff}\ and\ \citenamefont {Ziff}(2020)}]{Ziff}%
  \BibitemOpen
  \bibfield  {author} {\bibinfo {author} {\bibfnamefont {A.~L.}\ \bibnamefont
  {Ziff}}\ and\ \bibinfo {author} {\bibfnamefont {R.~M.}\ \bibnamefont
  {Ziff}},\ }\href {\doibase 10.1101/2020.02.16.2002382} {\bibfield  {journal}
  {\bibinfo  {journal} {MedRxiv preprint}\ } (\bibinfo {year} {2020}),\
  10.1101/2020.02.16.2002382}\BibitemShut {NoStop}%
\bibitem [{\citenamefont {Bianconi}\ and\ \citenamefont
  {Krapivsky}(2020)}]{bianconi2020epidemics}%
  \BibitemOpen
  \bibfield  {author} {\bibinfo {author} {\bibfnamefont {G.}~\bibnamefont
  {Bianconi}}\ and\ \bibinfo {author} {\bibfnamefont {P.~L.}\ \bibnamefont
  {Krapivsky}},\ }\href@noop {} {\bibfield  {journal} {\bibinfo  {journal}
  {arXiv preprint arXiv:2004.03934}\ } (\bibinfo {year} {2020})}\BibitemShut
  {NoStop}%
\bibitem [{\citenamefont {Nekovee}(2020)}]{Nekovee}%
  \BibitemOpen
  \bibfield  {author} {\bibinfo {author} {\bibfnamefont {M.}~\bibnamefont
  {Nekovee}},\ }\href {\doibase 10.1101/2020.05.18.20105445} {\bibfield
  {journal} {\bibinfo  {journal} {medRxiv}\ } (\bibinfo {year} {2020}),\
  10.1101/2020.05.18.20105445}\BibitemShut {NoStop}%
\bibitem [{\citenamefont {Valba}\ \emph {et~al.}(2020)\citenamefont {Valba},
  \citenamefont {Avetisov}, \citenamefont {Gorsky},\ and\ \citenamefont
  {Nechaev}}]{Nechaev}%
  \BibitemOpen
  \bibfield  {author} {\bibinfo {author} {\bibfnamefont {O.}~\bibnamefont
  {Valba}}, \bibinfo {author} {\bibfnamefont {V.}~\bibnamefont {Avetisov}},
  \bibinfo {author} {\bibfnamefont {A.}~\bibnamefont {Gorsky}}, \ and\ \bibinfo
  {author} {\bibfnamefont {S.}~\bibnamefont {Nechaev}},\ }\href@noop {}
  {\bibfield  {journal} {\bibinfo  {journal} {arXiv:2003.12290}\ } (\bibinfo
  {year} {2020})}\BibitemShut {NoStop}%
\bibitem [{\citenamefont {Blasius}(2020)}]{Blasius}%
  \BibitemOpen
  \bibfield  {author} {\bibinfo {author} {\bibfnamefont {B.}~\bibnamefont
  {Blasius}},\ }\href@noop {} {\bibfield  {journal} {\bibinfo  {journal}
  {arXiv:2004.00940}\ } (\bibinfo {year} {2020})}\BibitemShut {NoStop}%
\bibitem [{\citenamefont {Brandenburg}(2020)}]{quadratic}%
  \BibitemOpen
  \bibfield  {author} {\bibinfo {author} {\bibfnamefont {A.}~\bibnamefont
  {Brandenburg}},\ }\href@noop {} {\bibfield  {journal} {\bibinfo  {journal}
  {arXiv preprint arXiv:2002.03638}\ } (\bibinfo {year} {2020})}\BibitemShut
  {NoStop}%
\bibitem [{\citenamefont {Murray}(2007)}]{murray2007mathematical}%
  \BibitemOpen
  \bibfield  {author} {\bibinfo {author} {\bibfnamefont {J.~D.}\ \bibnamefont
  {Murray}},\ }\href@noop {} {\emph {\bibinfo {title} {Mathematical biology: I.
  An introduction}}},\ Vol.~\bibinfo {volume} {17}\ (\bibinfo  {publisher}
  {Springer Science \& Business Media New York},\ \bibinfo {year}
  {2007})\BibitemShut {NoStop}%
\bibitem [{\citenamefont {Krapivsky}\ \emph {et~al.}(2010)\citenamefont
  {Krapivsky}, \citenamefont {Redner},\ and\ \citenamefont
  {Ben-Naim}}]{krapivsky2010kinetic}%
  \BibitemOpen
  \bibfield  {author} {\bibinfo {author} {\bibfnamefont {P.~L.}\ \bibnamefont
  {Krapivsky}}, \bibinfo {author} {\bibfnamefont {S.}~\bibnamefont {Redner}}, \
  and\ \bibinfo {author} {\bibfnamefont {E.}~\bibnamefont {Ben-Naim}},\
  }\href@noop {} {\emph {\bibinfo {title} {A kinetic view of statistical
  physics}}}\ (\bibinfo  {publisher} {Cambridge University Press, Cambridge},\
  \bibinfo {year} {2010})\BibitemShut {NoStop}%
\bibitem [{\citenamefont {Barab{\'a}si}\ \emph {et~al.}(2016)\citenamefont
  {Barab{\'a}si} \emph {et~al.}}]{barabasi2016network}%
  \BibitemOpen
  \bibfield  {author} {\bibinfo {author} {\bibfnamefont {A.-L.}\ \bibnamefont
  {Barab{\'a}si}} \emph {et~al.},\ }\href@noop {} {\emph {\bibinfo {title}
  {Network science}}}\ (\bibinfo  {publisher} {Cambridge University Press,
  Cambridge},\ \bibinfo {year} {2016})\BibitemShut {NoStop}%
\bibitem [{\citenamefont {Newman}(2010)}]{newman2018networks}%
  \BibitemOpen
  \bibfield  {author} {\bibinfo {author} {\bibfnamefont {M.}~\bibnamefont
  {Newman}},\ }\href@noop {} {\emph {\bibinfo {title} {Networks}}}\ (\bibinfo
  {publisher} {Oxford University Press, Oxford},\ \bibinfo {year}
  {2010})\BibitemShut {NoStop}%
\bibitem [{\citenamefont {Bianconi}(2018)}]{bianconi2018multilayer}%
  \BibitemOpen
  \bibfield  {author} {\bibinfo {author} {\bibfnamefont {G.}~\bibnamefont
  {Bianconi}},\ }\href@noop {} {\emph {\bibinfo {title} {Multilayer networks:
  structure and function}}}\ (\bibinfo  {publisher} {Oxford University Press,
  Oxford},\ \bibinfo {year} {2018})\BibitemShut {NoStop}%
\bibitem [{\citenamefont {Barrat}\ \emph {et~al.}(2008)\citenamefont {Barrat},
  \citenamefont {Barthelemy},\ and\ \citenamefont
  {Vespignani}}]{barrat2008dynamical}%
  \BibitemOpen
  \bibfield  {author} {\bibinfo {author} {\bibfnamefont {A.}~\bibnamefont
  {Barrat}}, \bibinfo {author} {\bibfnamefont {M.}~\bibnamefont {Barthelemy}},
  \ and\ \bibinfo {author} {\bibfnamefont {A.}~\bibnamefont {Vespignani}},\
  }\href@noop {} {\emph {\bibinfo {title} {Dynamical processes on complex
  networks}}}\ (\bibinfo  {publisher} {Cambridge university press},\ \bibinfo
  {year} {2008})\BibitemShut {NoStop}%
\bibitem [{\citenamefont {Dorogovtsev}(2010)}]{dorogovtsev2010lectures}%
  \BibitemOpen
  \bibfield  {author} {\bibinfo {author} {\bibfnamefont {S.~N.}\ \bibnamefont
  {Dorogovtsev}},\ }\href@noop {} {\emph {\bibinfo {title} {Lectures on complex
  networks}}},\ Vol.~\bibinfo {volume} {24}\ (\bibinfo  {publisher} {Oxford
  University Press, Oxford},\ \bibinfo {year} {2010})\BibitemShut {NoStop}%
\bibitem [{\citenamefont {Pastor-Satorras}\ \emph {et~al.}(2015)\citenamefont
  {Pastor-Satorras}, \citenamefont {Castellano}, \citenamefont {Van~Mieghem},\
  and\ \citenamefont {Vespignani}}]{pastor2015epidemic}%
  \BibitemOpen
  \bibfield  {author} {\bibinfo {author} {\bibfnamefont {R.}~\bibnamefont
  {Pastor-Satorras}}, \bibinfo {author} {\bibfnamefont {C.}~\bibnamefont
  {Castellano}}, \bibinfo {author} {\bibfnamefont {P.}~\bibnamefont
  {Van~Mieghem}}, \ and\ \bibinfo {author} {\bibfnamefont {A.}~\bibnamefont
  {Vespignani}},\ }\href@noop {} {\bibfield  {journal} {\bibinfo  {journal}
  {Rev. Mod. Phys.}\ }\textbf {\bibinfo {volume} {87}},\ \bibinfo {pages} {925}
  (\bibinfo {year} {2015})}\BibitemShut {NoStop}%
\bibitem [{\citenamefont {Porter}\ and\ \citenamefont
  {Gleeson}(2016)}]{porter2016dynamical}%
  \BibitemOpen
  \bibfield  {author} {\bibinfo {author} {\bibfnamefont {M.~A.}\ \bibnamefont
  {Porter}}\ and\ \bibinfo {author} {\bibfnamefont {J.~P.}\ \bibnamefont
  {Gleeson}},\ }\href@noop {} {\bibfield  {journal} {\bibinfo  {journal}
  {Frontiers in Applied Dynamical Systems: Reviews and Tutorials}\ }\textbf
  {\bibinfo {volume} {4}} (\bibinfo {year} {2016})}\BibitemShut {NoStop}%
\bibitem [{\citenamefont {Nanni}\ \emph {et~al.}(2020)\citenamefont {Nanni},
  \citenamefont {Andrienko}, \citenamefont {Boldrini}, \citenamefont {Bonchi},
  \citenamefont {Cattuto}, \citenamefont {Chiaromonte}, \citenamefont
  {Comand{\'e}}, \citenamefont {Conti}, \citenamefont {Cot{\'e}}, \citenamefont
  {Dignum} \emph {et~al.}}]{nanni2020give}%
  \BibitemOpen
  \bibfield  {author} {\bibinfo {author} {\bibfnamefont {M.}~\bibnamefont
  {Nanni}}, \bibinfo {author} {\bibfnamefont {G.}~\bibnamefont {Andrienko}},
  \bibinfo {author} {\bibfnamefont {C.}~\bibnamefont {Boldrini}}, \bibinfo
  {author} {\bibfnamefont {F.}~\bibnamefont {Bonchi}}, \bibinfo {author}
  {\bibfnamefont {C.}~\bibnamefont {Cattuto}}, \bibinfo {author} {\bibfnamefont
  {F.}~\bibnamefont {Chiaromonte}}, \bibinfo {author} {\bibfnamefont
  {G.}~\bibnamefont {Comand{\'e}}}, \bibinfo {author} {\bibfnamefont
  {M.}~\bibnamefont {Conti}}, \bibinfo {author} {\bibfnamefont
  {M.}~\bibnamefont {Cot{\'e}}}, \bibinfo {author} {\bibfnamefont
  {F.}~\bibnamefont {Dignum}},  \emph {et~al.},\ }\href@noop {} {\bibfield
  {journal} {\bibinfo  {journal} {arXiv preprint arXiv:2004.05222}\ } (\bibinfo
  {year} {2020})}\BibitemShut {NoStop}%
\bibitem [{\citenamefont {Gross}\ \emph {et~al.}(2006)\citenamefont {Gross},
  \citenamefont {D’Lima},\ and\ \citenamefont {Blasius}}]{gross1}%
  \BibitemOpen
  \bibfield  {author} {\bibinfo {author} {\bibfnamefont {T.}~\bibnamefont
  {Gross}}, \bibinfo {author} {\bibfnamefont {C.~J.~D.}\ \bibnamefont
  {D’Lima}}, \ and\ \bibinfo {author} {\bibfnamefont {B.}~\bibnamefont
  {Blasius}},\ }\href@noop {} {\bibfield  {journal} {\bibinfo  {journal} {Phys.
  Rev. Lett.}\ }\textbf {\bibinfo {volume} {96}},\ \bibinfo {pages} {208701}
  (\bibinfo {year} {2006})}\BibitemShut {NoStop}%
\bibitem [{\citenamefont {Gross}\ and\ \citenamefont {Sayama}(2009)}]{gross2}%
  \BibitemOpen
  \bibfield  {author} {\bibinfo {author} {\bibfnamefont {T.}~\bibnamefont
  {Gross}}\ and\ \bibinfo {author} {\bibfnamefont {H.}~\bibnamefont {Sayama}},\
  }in\ \href@noop {} {\emph {\bibinfo {booktitle} {Adaptive networks}}}\
  (\bibinfo  {publisher} {Springer},\ \bibinfo {year} {2009})\ pp.\ \bibinfo
  {pages} {1--8}\BibitemShut {NoStop}%
\bibitem [{\citenamefont {Mora}\ and\ \citenamefont {Bialek}(2011)}]{Bialek}%
  \BibitemOpen
  \bibfield  {author} {\bibinfo {author} {\bibfnamefont {T.}~\bibnamefont
  {Mora}}\ and\ \bibinfo {author} {\bibfnamefont {W.}~\bibnamefont {Bialek}},\
  }\href@noop {} {\bibfield  {journal} {\bibinfo  {journal} {Journal of
  Statistical Physics}\ }\textbf {\bibinfo {volume} {144}},\ \bibinfo {pages}
  {268} (\bibinfo {year} {2011})}\BibitemShut {NoStop}%
\bibitem [{\citenamefont {Munoz}(2018)}]{Munoz}%
  \BibitemOpen
  \bibfield  {author} {\bibinfo {author} {\bibfnamefont {M.~A.}\ \bibnamefont
  {Munoz}},\ }\href@noop {} {\bibfield  {journal} {\bibinfo  {journal} {Reviews
  of Modern Physics}\ }\textbf {\bibinfo {volume} {90}},\ \bibinfo {pages}
  {031001} (\bibinfo {year} {2018})}\BibitemShut {NoStop}%
\bibitem [{\citenamefont {Gleeson}\ and\ \citenamefont
  {Durrett}(2017)}]{gleeson2017temporal}%
  \BibitemOpen
  \bibfield  {author} {\bibinfo {author} {\bibfnamefont {J.~P.}\ \bibnamefont
  {Gleeson}}\ and\ \bibinfo {author} {\bibfnamefont {R.}~\bibnamefont
  {Durrett}},\ }\href@noop {} {\bibfield  {journal} {\bibinfo  {journal}
  {Nature Communications}\ }\textbf {\bibinfo {volume} {8}},\ \bibinfo {pages}
  {1} (\bibinfo {year} {2017})}\BibitemShut {NoStop}%
\bibitem [{\citenamefont {Ben-Naim}\ and\ \citenamefont
  {Krapivsky}(2004)}]{uno}%
  \BibitemOpen
  \bibfield  {author} {\bibinfo {author} {\bibfnamefont {E.}~\bibnamefont
  {Ben-Naim}}\ and\ \bibinfo {author} {\bibfnamefont {P.~L.}\ \bibnamefont
  {Krapivsky}},\ }\href@noop {} {\bibfield  {journal} {\bibinfo  {journal}
  {Phys. Rev. E}\ }\textbf {\bibinfo {volume} {69}},\ \bibinfo {pages} {050901}
  (\bibinfo {year} {2004})}\BibitemShut {NoStop}%
\bibitem [{\citenamefont {Ben-Naim}\ and\ \citenamefont
  {Krapivsky}(2012)}]{due}%
  \BibitemOpen
  \bibfield  {author} {\bibinfo {author} {\bibfnamefont {E.}~\bibnamefont
  {Ben-Naim}}\ and\ \bibinfo {author} {\bibfnamefont {P.}~\bibnamefont
  {Krapivsky}},\ }\href@noop {} {\bibfield  {journal} {\bibinfo  {journal}
  {Eur. Phys. J. B}\ }\textbf {\bibinfo {volume} {85}},\ \bibinfo {pages} {145}
  (\bibinfo {year} {2012})}\BibitemShut {NoStop}%
\bibitem [{\citenamefont {Tom{\'e}}\ and\ \citenamefont
  {Ziff}(2010)}]{tome2010critical}%
  \BibitemOpen
  \bibfield  {author} {\bibinfo {author} {\bibfnamefont {T.}~\bibnamefont
  {Tom{\'e}}}\ and\ \bibinfo {author} {\bibfnamefont {R.~M.}\ \bibnamefont
  {Ziff}},\ }\href@noop {} {\bibfield  {journal} {\bibinfo  {journal} {Phys.
  Rev. E}\ }\textbf {\bibinfo {volume} {82}},\ \bibinfo {pages} {051921}
  (\bibinfo {year} {2010})}\BibitemShut {NoStop}%
\bibitem [{\citenamefont {Zapperi}\ \emph {et~al.}(1995)\citenamefont
  {Zapperi}, \citenamefont {Lauritsen},\ and\ \citenamefont
  {Stanley}}]{zapperi1}%
  \BibitemOpen
  \bibfield  {author} {\bibinfo {author} {\bibfnamefont {S.}~\bibnamefont
  {Zapperi}}, \bibinfo {author} {\bibfnamefont {K.~B.}\ \bibnamefont
  {Lauritsen}}, \ and\ \bibinfo {author} {\bibfnamefont {H.~E.}\ \bibnamefont
  {Stanley}},\ }\href@noop {} {\bibfield  {journal} {\bibinfo  {journal} {Phys.
  Rev. Lett.}\ }\textbf {\bibinfo {volume} {75}},\ \bibinfo {pages} {4071}
  (\bibinfo {year} {1995})}\BibitemShut {NoStop}%
\bibitem [{\citenamefont {Lauritsen}\ \emph {et~al.}(1996)\citenamefont
  {Lauritsen}, \citenamefont {Zapperi},\ and\ \citenamefont
  {Stanley}}]{zapperi2}%
  \BibitemOpen
  \bibfield  {author} {\bibinfo {author} {\bibfnamefont {K.~B.}\ \bibnamefont
  {Lauritsen}}, \bibinfo {author} {\bibfnamefont {S.}~\bibnamefont {Zapperi}},
  \ and\ \bibinfo {author} {\bibfnamefont {H.~E.}\ \bibnamefont {Stanley}},\
  }\href@noop {} {\bibfield  {journal} {\bibinfo  {journal} {Physical Review
  E}\ }\textbf {\bibinfo {volume} {54}},\ \bibinfo {pages} {2483} (\bibinfo
  {year} {1996})}\BibitemShut {NoStop}%
\bibitem [{\citenamefont {Henkel}\ \emph {et~al.}(2008)\citenamefont {Henkel},
  \citenamefont {Hinrichsen},\ and\ \citenamefont {L{\"u}beck}}]{Hinrichsen}%
  \BibitemOpen
  \bibfield  {author} {\bibinfo {author} {\bibfnamefont {M.}~\bibnamefont
  {Henkel}}, \bibinfo {author} {\bibfnamefont {H.}~\bibnamefont {Hinrichsen}},
  \ and\ \bibinfo {author} {\bibfnamefont {S.}~\bibnamefont {L{\"u}beck}},\
  }\href@noop {} {\emph {\bibinfo {title} {Non-equilibrium phase transitions:
  Absorbing Phase Transitions}}},\ Vol.~\bibinfo {volume} {1}\ (\bibinfo
  {publisher} {Springer},\ \bibinfo {year} {2008})\BibitemShut {NoStop}%
\bibitem [{\citenamefont {Janssen}\ \emph {et~al.}(1989)\citenamefont
  {Janssen}, \citenamefont {Schaub},\ and\ \citenamefont
  {Schmittmann}}]{slip_exponent}%
  \BibitemOpen
  \bibfield  {author} {\bibinfo {author} {\bibfnamefont {H.}~\bibnamefont
  {Janssen}}, \bibinfo {author} {\bibfnamefont {B.}~\bibnamefont {Schaub}}, \
  and\ \bibinfo {author} {\bibfnamefont {B.}~\bibnamefont {Schmittmann}},\
  }\href@noop {} {\bibfield  {journal} {\bibinfo  {journal} {Zeitschrift
  f{\"u}r Physik B Condensed Matter}\ }\textbf {\bibinfo {volume} {73}},\
  \bibinfo {pages} {539} (\bibinfo {year} {1989})}\BibitemShut {NoStop}%
\bibitem [{\citenamefont {Hinrichsen}\ and\ \citenamefont
  {{\'O}dor}(1998)}]{hinrichsen1998correlated}%
  \BibitemOpen
  \bibfield  {author} {\bibinfo {author} {\bibfnamefont {H.}~\bibnamefont
  {Hinrichsen}}\ and\ \bibinfo {author} {\bibfnamefont {G.}~\bibnamefont
  {{\'O}dor}},\ }\href@noop {} {\bibfield  {journal} {\bibinfo  {journal}
  {Physical Review E}\ }\textbf {\bibinfo {volume} {58}},\ \bibinfo {pages}
  {311} (\bibinfo {year} {1998})}\BibitemShut {NoStop}%
\bibitem [{\citenamefont {Radicchi}\ \emph {et~al.}(2020)\citenamefont
  {Radicchi}, \citenamefont {Castellano}, \citenamefont {Flammini},
  \citenamefont {Mu{\~n}oz},\ and\ \citenamefont
  {Notarmuzi}}]{radicchi2020classes}%
  \BibitemOpen
  \bibfield  {author} {\bibinfo {author} {\bibfnamefont {F.}~\bibnamefont
  {Radicchi}}, \bibinfo {author} {\bibfnamefont {C.}~\bibnamefont
  {Castellano}}, \bibinfo {author} {\bibfnamefont {A.}~\bibnamefont
  {Flammini}}, \bibinfo {author} {\bibfnamefont {M.~A.}\ \bibnamefont
  {Mu{\~n}oz}}, \ and\ \bibinfo {author} {\bibfnamefont {D.}~\bibnamefont
  {Notarmuzi}},\ }\href@noop {} {\bibfield  {journal} {\bibinfo  {journal}
  {Phys. Rev. Research}\ }\textbf {\bibinfo {volume} {2}},\ \bibinfo {pages}
  {033171} (\bibinfo {year} {2020})}\BibitemShut {NoStop}%
\bibitem [{\citenamefont {Goh}\ \emph {et~al.}(2003)\citenamefont {Goh},
  \citenamefont {Lee}, \citenamefont {Kahng},\ and\ \citenamefont {Kim}}]{Goh}%
  \BibitemOpen
  \bibfield  {author} {\bibinfo {author} {\bibfnamefont {K.-I.}\ \bibnamefont
  {Goh}}, \bibinfo {author} {\bibfnamefont {D.-S.}\ \bibnamefont {Lee}},
  \bibinfo {author} {\bibfnamefont {B.}~\bibnamefont {Kahng}}, \ and\ \bibinfo
  {author} {\bibfnamefont {D.}~\bibnamefont {Kim}},\ }\href@noop {} {\bibfield
  {journal} {\bibinfo  {journal} {Phys. Rev. Lett.}\ }\textbf {\bibinfo
  {volume} {91}},\ \bibinfo {pages} {148701} (\bibinfo {year}
  {2003})}\BibitemShut {NoStop}%
\bibitem [{\citenamefont {Barab{\'a}si}\ and\ \citenamefont
  {Stanley}(1995)}]{barabasi1995fractal}%
  \BibitemOpen
  \bibfield  {author} {\bibinfo {author} {\bibfnamefont {A.-L.}\ \bibnamefont
  {Barab{\'a}si}}\ and\ \bibinfo {author} {\bibfnamefont {H.~E.}\ \bibnamefont
  {Stanley}},\ }\href@noop {} {\emph {\bibinfo {title} {Fractal concepts in
  surface growth}}}\ (\bibinfo  {publisher} {Cambridge university press},\
  \bibinfo {year} {1995})\BibitemShut {NoStop}%
\bibitem [{\citenamefont {Marro}\ and\ \citenamefont {Dickman}(2005)}]{Marro}%
  \BibitemOpen
  \bibfield  {author} {\bibinfo {author} {\bibfnamefont {J.}~\bibnamefont
  {Marro}}\ and\ \bibinfo {author} {\bibfnamefont {R.}~\bibnamefont
  {Dickman}},\ }\href@noop {} {\emph {\bibinfo {title} {Nonequilibrium phase
  transitions in lattice models}}}\ (\bibinfo  {publisher} {Cambridge
  University Press},\ \bibinfo {year} {2005})\BibitemShut {NoStop}%
\bibitem [{\citenamefont {Krapivsky}(2020)}]{krapivsky2020infection}%
  \BibitemOpen
  \bibfield  {author} {\bibinfo {author} {\bibfnamefont {P.}~\bibnamefont
  {Krapivsky}},\ }\href@noop {} {\bibfield  {journal} {\bibinfo  {journal}
  {arXiv preprint arXiv:2009.08940}\ } (\bibinfo {year} {2020})}\BibitemShut
  {NoStop}%
\bibitem [{\citenamefont {Gillespie}(1976)}]{gillespie1976general}%
  \BibitemOpen
  \bibfield  {author} {\bibinfo {author} {\bibfnamefont {D.~T.}\ \bibnamefont
  {Gillespie}},\ }\href@noop {} {\bibfield  {journal} {\bibinfo  {journal}
  {Journal of Computational Physics}\ }\textbf {\bibinfo {volume} {22}},\
  \bibinfo {pages} {403} (\bibinfo {year} {1976})}\BibitemShut {NoStop}%
\end{thebibliography}%
\appendix
\section{Stochastic SIR dynamics on well mixed populations}
The critical SIR dynamics in a well-mixed population of $N$ individuals is simulated with the following implementation of the Gillespie algorithm~\cite{gillespie1976general}. We indicate with $S(t),I(t)$ and $R(t)$ respectively the number of susceptible, infected and removed individuals as a function of time $t$.
We start from the initial condition of $I(0)=n_0$,  $S(0)=N-n_0$ and $R(0)=0$.
At each elementary step,  the algorithm proceeds as follows:
\begin{itemize}
\item[(i)]
Time increases by the amount $\Delta t$
\bea
t\to t+\Delta t,
\eea
where $\Delta t$ is given by 
\bea
\Delta t=\frac{-\log(q)}{\lambda S(t)I(t)+I(t)}.
\eea
with $q \sim \textrm{Unif.(0,1)}$, i.e., a random variate extracted from the uniform distribution in the domain $(0, 1)$.
\item[(ii)]
With probability 
\bea
p=\frac{\lambda S(t)I(t)}{\lambda S(t)I(t)+I(t)}
\eea
a susceptible individual becomes infected, i.e.,
\bea
S(t+\Delta t)&=&S(t)-1,\nonumber \\
I(t+\Delta t)&=& I(t)+1.
\eea
\item[(iii)]
With probability $1-p$ an infected individual is removed, i.e.,
\bea
I(t+\Delta t)&=&I(t)-1,\nonumber \\
R(t+\Delta t)&=&R(t)+1.
\eea
\end{itemize}
The critical dynamics is obtained by setting $\lambda=1$. 
The steps of the algorithms are iterated until the number of  infected individuals is zero. This happens at time $T$, i.e., the duration of the outbreak.
The size of the outbreak is given by $R(T)$.
\end{document}